\documentclass[aps,pre,preprint,,superscriptaddress,showpacs,nofootinbib]{revtex4-1}

\usepackage[utf8]{inputenc}
\usepackage[english]{babel}

\usepackage{hyperref} 
\hypersetup{colorlinks=true,       
urlcolor=blue,          
linkcolor=blue,         
citecolor=blue,         
filecolor=blue,          
urlcolor=blue}

\usepackage{amssymb,amsmath}
\usepackage{graphicx}

\usepackage{setspace}

\begin{document}

\ \\\textbf{ACCEPTED MANUSCRIPT}\vspace{2.5mm}\\ 
Published in final edited form as:\\
Chella, F., Pizzella, V., Zappasodi, F., Nolte, G., Marzetti, L. (2016). Bispectral pairwise interacting source analysis for identifying systems of cross-frequency interacting brain sources from electroencephalographic or magnetoencephalographic signals. \textit{Physical Review E}, 93, 052420. DOI: \href{https://doi.org/10.1103/PhysRevE.93.052420}{https://doi.org/10.1103/PhysRevE.93.052420}\vspace{2.5mm}\\ 
Copyright information and permissions: \\
\href{https://journals.aps.org/authors/transfer-of-copyright-agreement}{https://journals.aps.org/authors/transfer-of-copyright-agreement} 
\vspace{0.5cm}


\title{Bispectral pairwise interacting source analysis for identifying systems of cross-frequency interacting brain sources from electroencephalographic or magnetoencephalographic signals}



\author{Federico Chella}
\email[E-mail address: ]{f.chella@unich.it}
\affiliation{Department of Neuroscience, Imaging and Clinical Sciences, ``G. d'Annunzio" University of Chieti-Pescara, via dei Vestini 31, 66100 Chieti, Italy}
\affiliation{Institute for Advanced Biomedical Technologies, ``G. d'Annunzio" University of Chieti-Pescara, via dei Vestini 31, 66100 Chieti, Italy}

\author{Vittorio Pizzella}
\affiliation{Department of Neuroscience, Imaging and Clinical Sciences, ``G. d'Annunzio" University of Chieti-Pescara, via dei Vestini 31, 66100 Chieti, Italy}
\affiliation{Institute for Advanced Biomedical Technologies, ``G. d'Annunzio" University of Chieti-Pescara, via dei Vestini 31, 66100 Chieti, Italy}

\author{Filippo Zappasodi}
\affiliation{Department of Neuroscience, Imaging and Clinical Sciences, ``G. d'Annunzio" University of Chieti-Pescara, via dei Vestini 31, 66100 Chieti, Italy}
\affiliation{Institute for Advanced Biomedical Technologies, ``G. d'Annunzio" University of Chieti-Pescara, via dei Vestini 31, 66100 Chieti, Italy}

\author{Guido Nolte}
\affiliation{Department of Neurophysiology and Pathophysiology, University Medical Center Hamburg-Eppendorf, Martinistra\ss e 52, D-20246 Hamburg, Germany}

\author{Laura Marzetti}
\affiliation{Department of Neuroscience, Imaging and Clinical Sciences, ``G. d'Annunzio" University of Chieti-Pescara, via dei Vestini 31, 66100 Chieti, Italy}
\affiliation{Institute for Advanced Biomedical Technologies, ``G. d'Annunzio" University of Chieti-Pescara, via dei Vestini 31, 66100 Chieti, Italy}



\begin{abstract}
Brain cognitive functions arise through the coordinated activity of several brain regions, which actually form complex dynamical systems operating at multiple frequencies. These systems often consist of interacting subsystems, whose characterization is of importance for a complete understanding of the brain interaction processes. To address this issue, we present a technique, namely the bispectral Pairwise Interacting Source Analysis (biPISA), for analyzing systems of cross-frequency interacting brain sources when multichannel electroencephalographic (EEG) or magnetoencephalographic (MEG) data are available. Specifically, the biPISA allows to identify one or many subsystems of cross-frequency interacting sources by decomposing the antisymmetric components of the cross-bispectra between EEG or MEG signals, based on the assumption that interactions are pairwise. Thanks to the properties of the antisymmetric components of the cross-bispectra, biPISA is also robust to spurious interactions arising from mixing artifacts, i.e. volume conduction or field spread, which always affect EEG or MEG functional connectivity estimates. This method is an extension of the Pairwise Interacting Source Analysis (PISA), which was originally introduced for investigating interactions at the same frequency, to the study of cross-frequency interactions. The effectiveness of this approach is demonstrated in simulations for up to three interacting source pairs, and for real MEG recordings of spontaneous brain activity. Simulations show that the performances of biPISA in estimating the phase difference between the interacting sources are affected by the increasing level of noise rather than by the number of the interacting subsystems. The analysis of real MEG data reveals an interaction between two pairs of sources of central mu and beta rhythms, localizing in the proximity of the left and right central sulci.
\end{abstract}


\maketitle

\section{Introduction}
Synchronization is a phenomenon which is ubiquitous in nature, playing a fundamental role in many different branches of science such as physics, chemistry, biology, engineering and mechanics, medicine and life sciences, ecology, sociology, or even in fine arts \cite{pikovsky03,osipov07,arenas08}. Synchronization is possible if at least two elements are coupled, but it much more often involves multiple, even thousands of subsystems that interact with each other, typically in a nonlinear fashion. Increasingly the characterization of the collective behavior displayed by interacting components is of importance for understanding and ultimately designing systems \cite{strogatz01,reka02,boccaletti06}. 

A striking example of such large and complex systems with synchronous dynamical components is the human brain. Indeed, phase synchronization of oscillatory brain activity has been recognized to play a central role in neuronal communication both at local and large scale, possibly serving as a mechanism to regulate the integration and flow of cognitive contents on fast timescales relevant to behavior \cite[e.g.][]{varela01,fries05,fries09,fell11,engel13}. Indeed, phase coupling has been observed in multiple defined frequency bands, i.e. from $\sim1$ Hz to 150 Hz, and has been shown to feature strong condition specific modulations while being less constrained by structural coupling \cite{engel13}. Furthermore, high frequency coupling (e.g. gamma band range) seems to regulate local synchronization within neuronal assemblies, whereas the interplay between different neuronal pools is served by the synchronization at lower frequency ranges \cite{vonstein00}. In addition, the building blocks defined by distinct frequencies can give rise to a more sophisticated coupling structure through the interactions between different frequencies, i.e. cross-frequency phase synchronization. This type of interaction has been shown to serve as carrier mechanism for the integration of spectrally distributed processing \cite{palva05,jensen07,canolty10,jirsa13}, providing a plausible physiological mechanism for linking activity that occurs at different temporal rates. 

In the above context, the development of methods for the detection of phase synchronization from the electrophysiological correlates of neuronal activity measured by noninvasive techniques, such as electroencephalography (EEG) or magnetoencephalography (MEG), plays a key role for investigating brain cognition and behavior. While the majority of methods developed so far have focused on the synchronization between neuronal oscillations at the same frequency or within the same frequency band \cite[e.g.][]{nunez97,nolte04,fries05,gross06,srinivasan07,siegel08,nolte10,marzetti13}, in the recent years, an increasing interest has been devoted to the development of methods for the detection of cross-frequency synchronization, including measures for the estimation of the $\mbox{\textit{n}:\textit{m}}$ phase-locking \cite{tass03,nikulin06,wendling09} and bispectral measures \cite{helbig06,schwilden06,wang07,darvas09a,darvas09b,jirsa13}. This paper aims to contribute in the latter direction, addressing methodological concerns which are often overlooked when estimating cross-frequency couplings using EEG or MEG. The major challenge is that the data are largely unknown mixture of the activities of brain sources and thus it is of fundamental importance to separate genuine interactions from mixing artifacts \cite{nunez97,srinivasan07,winter07,schoffelen09}. Indeed, MEG and EEG sensor level interaction might be severely biased by mixing effects \cite{nolte04,marzetti07} which artificially enhance the degree of coupling between channels independently of the actual interactions between brain sources. Although source localization methods may attenuate these effects, it is important to note that the unmixing of the sources is never perfect, and thus mixing artifacts are never completely abolished, even in the source space \cite{schoffelen09,sekihara11}.

To deal with the problem of mixing artifacts in relation to the use of bispectral measures, in \citet{chella14} we suggested to use the antisymmetric components of cross-bispectra between sensor recordings, inasmuch as these quantities cannot be generated by the superposition of independent sources and, thus, necessarily reflect genuine cross-frequency coupling. In that paper, we also proposed a fit based procedure for identifying the sources of the observed antisymmetric components of cross-bispectra, relying on an interaction model consisting of two neuronal sources. Despite the promising results obtained with the two-source model, the brain interaction dynamic requires, in general, more elaborated models. Indeed, the brain cognitive functions arise through the concerted activity of multiple brain regions, which actually form complex dynamical systems. Often, these systems consist of interacting subsystems, whose characterization is of importance for a complete understanding of the interaction processes.\\
To address this issue, in this paper we extend the Pairwise Interacting Source Analysis (PISA), originally developed by \citet{nolte06} with the aim of estimating multiple sources interacting at a given frequency, to the decomposition of cross-frequency interactions as observed by the antisymmetric components of the cross-bispectra. 

The paper is organized as follows. Section \ref{method_sec} includes the method description. Specifically, subsections \ref{back_sec} and \ref{norm_sec} introduce the definition and properties of the antisymmetric components of cross-bispectra, and subsection \ref{theo_sec} presents the theory for the extension of the Pairwise Interacting Source Analysis to the antisymmetric components of cross-bispectra (namely the biPISA approach). Section \ref{results_sec} describes the simulation-based assessment of the proposed method and an example of application to the analysis of real MEG data. A general discussion on the method and the results presented in this study is given in section \ref{discuss_subsec}. Final conclusions are drawn in section \ref{conclusion_sec}.
  
\section{Methods\label{method_sec}}
The biPISA approach is an extension of the Pairwise Interacting Source Analysis (PISA) \cite{nolte06} to the study of cross-frequency brain interactions through bispectral analysis of EEG or MEG signals. In analogy to PISA, biPISA allows to identify systems of interacting brain sources under the following assumptions: (i) the interactions are pairwise; and (ii) the number of interacting sources is not greater than the number of EEG or MEG recording channels. In biPISA, these two basic assumptions lead to a special model for the antisymmetric components of the cross-bispectrum in the same way in which they do, in PISA, for the imaginary part of the cross-spectrum. Being related to statistics of different orders, namely the biPISA to cross-bispectra (3$rd$ order) and the PISA to cross-spectra (2$nd$ order), the two methods investigate different types of phase synchronization in brain oscillatory activity: biPISA is sensitive to the synchronization of the phases of oscillatory components at different frequencies, while PISA is sensitive to the synchronization of the phases of oscillatory components at the same frequency in each interacting system. For the reader interested in technical details of PISA we refer to the original publication \cite{nolte06}, whereas the theory for biPISA is presented below.

\subsection{Theoretical background for antisymmetric bispectral measures\label{back_sec}}
We start by recalling some basic definitions and properties of bispectral analysis. Given the timeseries recorded at three EEG or MEG channels, say $i$, $j$ and $k$, without loss of generality assumed to be zero-mean, and denoting by $x_i(f)$, $x_j(f)$ and $x_k(f)$ their complex-valued Fourier transforms at frequency $f$, the cross-bispectrum can be estimated as \cite{nikias93}
\begin{equation}
B_{ijk}(f_1,f_2) = \left<x_i(f_1)x_j(f_2)x^*_k(f_1+f_2)\right>
\label{bispecdef}
\end{equation}
where $^*$ means the complex conjugation and $\left<\cdot\right>$ denotes taking the expectation value, i.e. the average over a sufficiently large number of signal realizations (or segments). The frequency of $x^*_k$ was set to  $f_1+f_2$ because all other choices lead to vanishing bispectra for spontaneous or task related data as it will be shown in appendix \ref{AppA}. The cross-bispectrum is a measure of the synchronization between the phases in channels $i$ and $j$ at two possibly different frequencies, $\phi_i(f_1)$ and $\phi_j(f_2)$, with respect to the phase in channel $k$ at a third frequency which is the sum of the other two, $\phi_k(f_3)$, such that $f_3=f_1+f_2$. Synchronization essentially means the coordination of phases in such a way that the generalized phase difference $\phi_i(f_1) + \phi_j(f_2) - \phi_k(f_3)$ stays close to a constant value. Such a phenomenon is called quadratic phase coupling \cite{kim78,nikias93} and it is conceptually different from the \textit{n}:\textit{m} phase locking which generally indicates the phase locking on \textit{n} cycles of one oscillation to \textit{m} cycles of another oscillation \cite{rosenblum96,tass98}. Indeed, the quadratic phase coupling requires, in the most general case, the interplay between three frequency components which, taken in pairs, might be not synchronous in the sense of the \textit{n}:\textit{m} phase locking. There is one case, however, in which the two phenomena match, and which we will see to be relevant in actual analysis. This is the case of $f_1=f_2=:f$ and $f_3=2f$, in which the quadratic phase coupling involves only two frequency components, i.e. one frequency and its double, thus matching the 1:2 phase locking.

In a previous work \cite{chella14}, we argued that the antisymmetric component of the cross-bispectrum\footnote{As in our previous work, we denote the antisymmetrizing operation on cross-bispectrum indices by a bracket notation in which $[\cdot]$ indicates antisymmetrization over subset of indices included in the brackets, e.g., $B_{i[jk]}= B_{ijk} - B_{ikj}$. In the event that indices to be antisymmetrized are not adjacent to each other, as in equation \ref{acb}, the preceding notation is extended by using vertical lines to exclude indices from the antisymmetrization, i.e.: $B_{[i|j|k]}= B_{ijk} - B_{kji}$.}
\begin{equation}
B_{[i|j|k]}(f_1,f_2)  =  B_{ijk}(f_1,f_2) - B_{kji}(f_1,f_2)
\label{acb}
\end{equation}
namely the difference between two cross-bispectra for which two channel indices have been switched, i.e. $i$ and $k$ in the above equation, has the advantage over the conventional cross-bispectrum to be not affected by the activity of independent noisy sources. Hence, the analysis of the antisymmetric component of the cross-bispectrym, $B_{[i|j|k]}$, rather than the conventional cross-bispectrum, $B_{ijk}$, is more suitable to study brain interactions. We will shortly rederive this result for the sake of completeness. We make the usual assumptions that the data have zero mean (or that the mean has been subtracted from the raw data), and that the observed signals $x_i(f)$ result from a linear superposition of the source signals $s_m(f)$, i.e.,
\begin{equation}
x_i(f) = \sum_{m} a_{im}s_m(f)
\label{linsup}
\end{equation}
with $a_{im}$ being real-valued coefficients, independent of the frequency, corresponding to the forward mapping of the $m$th source to the $i$th channel. We emphasize that, while in general $x_i(f)$ and $s_m(f)$ are complex-valued, the $a_{im}$ coefficients are real-valued, which is a consequence of the fact that, under the quasi-static approximation for the electromagnetic field, the signal propagation from sources to channels does not introduce observable phase distortions \cite{stinstra98}. We further assume that all sources are independent, i.e. there is no interaction between different sources, and insert that into equation \ref{bispecdef}
\begin{equation}
B_{ijk}(f_1,f_2) = \sum_m a_{im}a_{jm}a_{km}\left<s_m(f_1)s_m(f_2)s^*_m(f_1+f_2)\right>+coupling\ terms
\label{bispec_proof1R}
\end{equation}
The summation in the right-hand side of the above equation contains terms which reflect the interaction of each source with itself. On the contrary, the 'coupling terms' reflect the interaction between different sources, and contain expressions of the form $\left<s_m(f_1)s_n(f_2)s^*_p(f_1+f_2)\right>$ where not all indices $m,n,p$ are identical, i.e. at least one of the indices is different from the other two. If, e.g., this index is the first one, $m$, and all sources are independent, this term vanishes 
\begin{equation}
\left<s_m(f_1)s_n(f_2)s^*_p(f_1+f_2)\right>=\left<s_m(f_1)\right>\left<s_n(f_2)s^*_p(f_1+f_2)\right>=0
\end{equation}
and likewise for any other of the indices. Hence, for independent sources we get 
\begin{equation}
B_{ijk}(f_1,f_2) = \sum_m a_{im}a_{jm}a_{km}\left<s_m(f_1)s_m(f_2)s^*_m(f_1+f_2)\right>
\label{bispec_proof2}
\end{equation}
which is totally symmetric with respect to the three channel indices. From this, it follows that an antisymmetric combination with respect to any pair of indices must vanish for independent sources, whereas, if not-vanishing, it must necessarily reflect the presence of an interaction between different sources.

A general expression of the antisymmetric component of the cross-bispectrum between channels in terms of brain source activities is obtained by inserting equation \ref{linsup} in \ref{acb}, yielding to
\begin{equation}
B_{[i|j|k]}(f_1,f_2) = \sum_{m,n,p} a_{im}\,a_{jn}\,a_{kp}\,\mathcal{B}_{[m|n|p]}(f_1,f_2) 
\label{gen_acb}
\end{equation}
with the coupling terms
\begin{eqnarray}
 \mathcal{B}_{[m|n|p]}(f_1,f_2) &=& \left<s_m(f_1)s_n(f_2)s_p^*(f_1+f_2)\right> \nonumber\\ 
&&\;\; -\left<s_p(f_1)s_n(f_2)s_m^*(f_1+f_2)\right>
\label{acbso}
\end{eqnarray}
being the antisymmetric components of the cross-bispectra between sources (here and in the following, the indices $i$, $j$ and $k$ run over channels and the indices $m$, $n$ and $p$ run over sources).

\subsection{Normalization of antisymmetric bispectral measures\label{norm_sec}}
In analogy to cross-spectra, bispectral estimates depend on the signal amplitudes at the specific frequencies at which they are calculated. In order to assess whether the coupling between signals is high or low irrespective of their amplitudes, it is then necessary to normalize the bispectral values by a measure of signal strength. For the conventional cross-bispectrum, this is achieved by means of the bicoherence, i.e. a normalized version of the cross-bispectrum, which is the analogous of the coherence for the cross-spectrum. Different expressions for bicoherence have been suggested so far \cite[e.g.][]{brillinger65,kim79,hinich05,helbig06}, which essentially differ by the normalization factor adopted. \citet{shahbazi14} recently introduced a new normalization factor for the cross-bispectrum, which reads
\begin{equation}
N_{ijk}(f_1,f_2) = Q_i(f_1)Q_j(f_1)Q_k(f_1+f_2)
\label{uninorm}
\end{equation}
where
\begin{equation}
Q_i(f) = \left(\frac{1}{N_{s}}\sum_{n_{s}} \left|x_i(f,n_{s})\right|^3\right)^{1/3}
\end{equation}
with $x_i(f,n_{s})$ being the Fourier transform of channel $i$ at frequency $f$, for the $n_{s}$th of the $N_{s}$ segments into which the data are divided to estimate the cross-bispectra. This normalization factor is called univariate in the sense that it normalizes the cross-bispectrum by the signal amplitude at each channel separately, with the result that it does not depend on the interactions between channels. Moreover, it has the advantage over other existent normalizations that the absolute value of bicoherence is bounded by one, i.e.
\begin{equation}
\left|b_{ijk}\right|=\left|\frac{B_{ijk}}{N_{ijk}}\right| \leq 1
\label{uninormbound}
\end{equation}

Here, to the purpose of normalizing the antisymmetric components of the cross-bispectrum rather than the full cross-bispectrum, we define a slightly different normalization factor by taking the symmetric component of the univariate normalization factor, i.e.
\begin{equation}
N_{(i|j|k)}(f_1,f_2) = N_{ijk}(f_1,f_2) + N_{kji}(f_1,f_2)
\end{equation}
where the round bracket notation in the subscript, i.e. $(i|j|k)$, has been used in the above equation to denote the symmetrization operation over channel indices in the same way as the square bracket notation, i.e. $[i|j|k]$, was used to define the antisymmetrization operation. Accordingly, our bicoherence reads as follows:
\begin{equation}
b_{ijk}(f_1,f_2) =\frac{B_{ijk}(f_1,f_2)}{N_{(i|j|k)}(f_1,f_2)}
\end{equation}
The advantage of taking the symmetric component of $N_{ijk}$ at the denominator of the bicoherence is that the antisymmetric component of the bicoherence reads as the desired normalized version of the antisymmetric component of the cross-bispectrum, i.e.
\begin{equation}
b_{[i|j|k]}(f_1,f_2) =\frac{B_{[i|j|k]}(f_1,f_2)}{N_{(i|j|k)}(f_1,f_2)}
\label{acb_norm}
\end{equation}
whose magnitude is still bounded by one (see appendix \ref{AppB} for a proof).

\subsection{The biPISA approach\label{theo_sec}}
\subsubsection{Problem formulation}
The aim of biPISA is to identify the interacting brain sources from given estimates of the antisymmetric components of the cross-bispectra between channels. In practice, this means to find the $a_{im}$ coefficients in equation \ref{gen_acb} from the observed $B_{[i|j|k]}$, and to subsequently interpret them in terms of the field patterns of the interacting sources. 
The key assumption we make in the following is that interactions are pairwise, namely the interacting sources can be broken into a set of independent subsystems, each of them consisting of two interacting sources. Subsystem independence means that sources belonging to different subsystems do not interact with each other. Then, most of the coupling terms in equation \ref{gen_acb} vanish, leaving only the terms involving sources within the same pair. If we also denote by $s_{1q}$ and $s_{2q}$ the sources which form the $q$-th interacting pair and by $a_{iq}$ and $b_{iq}$ their respective coefficients at channel $i$, then equation \ref{gen_acb} can be rewritten as the sum of terms due to individual interacting subsystems
\begin{eqnarray}
B_{[i|j|k]}(f_1,f_2)  &=& \sum_{q} \biggl\{\Bigr(a_{iq}\,a_{jq}\,b_{kq}-a_{kq}\,a_{jq}\,b_{iq}\Bigr)  \alpha_q(f_1,f_2) \nonumber \\
  &&\!\!\!\!\!\!\!\!\! + \Bigr(a_{iq}\,b_{jq}\,b_{kq}-a_{kq}\,b_{jq}\,b_{iq}\Bigr)  \beta_q(f_1,f_2)\biggl\}
\label{pisamodel}
\end{eqnarray}
with
\begin{eqnarray}
\alpha_q(f_1,f_2)&=&\left<\right.s_{1q}(f_1)s_{1q}(f_2)s_{2q}^*(f_1+f_2)\left.\right> \nonumber\\
&& - \left<\right.s_{2q}(f_1) s_{1q}(f_2)s_{1q}^*(f_1+f_2)\left.\right> \label{alphaq_pisa}\\
\beta_q(f_1,f_2)&=&\left<\right.s_{1q}(f_1)s_{2q}(f_2)s_{2q}^*(f_1+f_2)\left.\right> \nonumber\\
&& - \left<\right.s_{2q}(f_1) s_{2q}(f_2) s_{1q}^*(f_1+f_2)\left.\right> 
\label{betaq_pisa}
\end{eqnarray}

\begin{figure*}[!ht] 
\centering
\includegraphics[width=16cm]{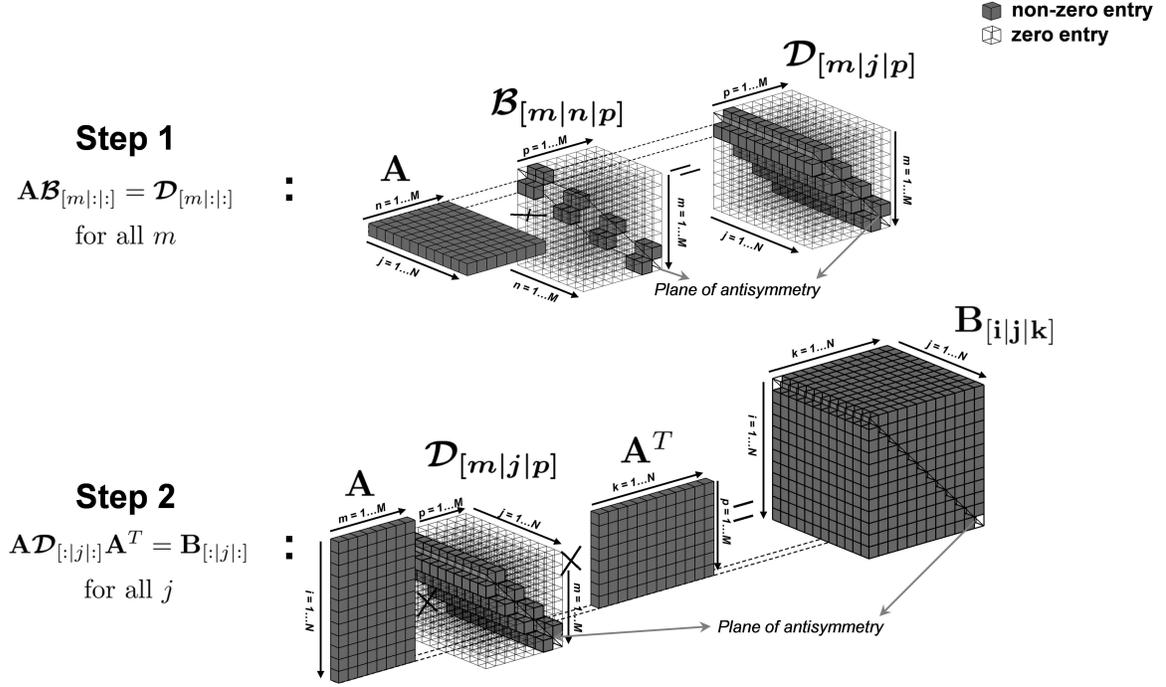}
\caption{\small{The two-step computation of the antisymmetric bispectral tensor between channel signals, $\mathbf{{B}_{[i|j|k]}}$, from the antisymmetric bispectral tensor between source signals, $\boldsymbol{\mathcal{B}_{[m|n|p]}}$}\label{tens_mult}}
\end{figure*}

The main point here is that the pairwise interaction assumption has led to a model for the antisymmetric components of the cross-bispectrum which can be efficiently solved in terms of source coefficients $a_{iq}$ and $b_{iq}$ by means of linear algebraic techniques.\\
To see this, we must first rewrite the equation \ref{pisamodel} by using a tensor representation. Let us, then, denote by $N$ be the number of EEG or MEG recording channels, by $M$ the (unknown) number of pairwise interacting sources and by $Q=M/2$ the number of the corresponding interacting pairs. Accordingly, $\mathbf{{B}_{[i|j|k]}}(f_1,f_2)$ is the three-way $N$$\times$$N$$\times$$N$ tensor collecting the antisymmetric components of the cross-bispectrum between channel signals at frequency pair $(f_1,f_2)$, and $\boldsymbol{\mathcal{B}_{[m|n|p]}}(f_1,f_2)$ is the three-way $M$$\times$$M$$\times$$M$ tensor collecting the antisymmetric components of the cross-bispectrum between source signals. The subscripts $\mathbf{[i|j|k]}$ and $\boldsymbol{[m|n|p]}$ are introduced to remind us that such tensors are antisymmetric in the $ik$ and $mp$ indices, respectively. Similarly, the $a_{im}$ coefficients are collected in an $N$$\times$$M$ matrix $\mathbf{A}$, whose $M$ columns read as the $N$-length vector topographies of sources into channels. Where necessary, the $N$-length vector topographies of sources belonging to the $q$th interacting pair will be denoted as $\mathbf{a}_q$ and $\mathbf{b}_q$. For the ease of reading, the tensor dependence on the frequency is omitted in the following.\\
The pairwise interaction assumption implies a special structure for the source-level tensor $\boldsymbol{\mathcal{B}_{[m|n|p]}}$. Indeed, without loss of generality, we can always arrange the interacting sources such that the $q$th pair is composed by consecutive sources , i.e. $s_1$ is interacting with $s_2$, $s_3$ with $s_4$, and so on. Then, it turns out that $\boldsymbol{\mathcal{B}_{[m|n|p]}}$ is a 2$\times$2$\times$2 block diagonal tensor, i.e., its non-zero entries fill $Q$ blocks of size 2$\times$2$\times$2 on the main diagonal, while all the entries outside these blocks, i.e. corresponding to cross-bispectra between sources which do not belong to the same interacting pair, vanish. Moreover, since the whole tensor is antisymmetric in the $mp$ indices, the entries on the diagonal slice $m=p$ are zero and, thus, each 2$\times$2$\times$2 block is uniquely defined by four complex values, which read $\pm\alpha_q$ and $\pm\beta_q$ for the $q$th block.\\
Using tensor representation, we can now rethink equation \ref{pisamodel} as a spatial transformation from $\boldsymbol{\mathcal{B}_{[m|n|p]}}$ to $\mathbf{{B}_{[i|j|k]}}$ according to the following two steps: (i) we first compute an $M$$\times$$N$$\times$$M$ intermediate tensor $\boldsymbol{\mathcal{D}_{[m|j|p]}}$ by left-multiplying by $\mathbf{A}$ each $m$-th slice of $\boldsymbol{\mathcal{B}_{[m|n|p]}}$; note that each $j$th slice of $\boldsymbol{\mathcal{D}_{[m|j|p]}}$ is a 2$\times$2 block diagonal matrix, as well as antisymmetric, which follows by construction since $\boldsymbol{\mathcal{B}_{[m|n|p]}}$ is an antisymmetric 2$\times$2$\times$2 block diagonal tensor; and (ii) each $j$-th slice of $\boldsymbol{\mathcal{D}_{[m|j|p]}}$ is left-multiplied by $\mathbf{A}$ and right-multiplied by $\mathbf{A}^T$ in order to obtain $\mathbf{{B}_{[i|j|k]}}$. These two steps are schematically depicted in figure \ref{tens_mult}.
\\
The main advantage of tensor representation is that it makes visible the inherent structure of bispectral data when pairwise interaction is assumed: each $j$th slice of $\mathbf{{B}_{[i|j|k]}}$ results from the mixing of an antisymmetric 2$\times$2 block diagonal matrix, i.e. the corresponding $j$th slice of $\boldsymbol{\mathcal{D}_{[m|j|p]}}$, through the coefficient matrix $\mathbf{A}$. We emphasize this result as we will now turn it upside down to the aim of estimating the matrix $\mathbf{A}$ of source coefficients.

For the ease of notation, let us denote by $\boldsymbol{\mathcal{D}}_{[:|j|:]}$ and $\mathbf{{B}}_{[:|j|:]}$ the $j$th slices of $\boldsymbol{\mathcal{D}_{[m|j|p]}}$ and $\mathbf{{B}_{[i|j|k]}}$, respectively. The above argument implies that a real valued demixing matrix $\mathbf{W}_1=\mathbf{A}^{-1}$ exists such that all the $j$th slices $\mathbf{{B}}_{[:|j|:]}$ are block-diagonalized, i.e. are transformed in 2$\times$2 block diagonal matrices $\boldsymbol{\mathcal{D}}_{[:|j|:]}$, according to
\begin{equation}
\boldsymbol{\mathcal{D}}_{[:|j|:]} = \mathbf{W}_1 \mathbf{{B}}_{[:|j|:]} \mathbf{W}_1^{\dagger}
\end{equation}
for $j=1...N$, with $\dagger$ denoting the complex conjugate transpose. Thus, we can estimate the demixing matrix $\mathbf{W}_1$ (and, accordingly, its inverse $\mathbf{A}$) by using the techniques for approximate joint block-diagonalization of a set of matrices. In particular, since all the $j$th slices of $\mathbf{{B}_{[i|j|k]}}$ are complex-valued matrices and, in general, their joint block-diagonalization will lead to a solution in the complex-domain, we further constrain $\mathbf{W}_1$ to be real-valued by requiring the joint block-diagonalization of both the real and imaginary parts of these slices.\\
In summary, the problem of finding the matrix of interacting source coefficients has been reduced to the joint block-diagonalization of a set of $2N$ real-valued antisymmetric matrices, namely $N$ from $\operatorname{Re}\left\{\mathbf{B}_{[:|j|:]}\right\}$ and $N$ from $\operatorname{Im}\left\{\mathbf{B}_{[:|j|:]}\right\}$, for $j=1...N$. In practice, these matrices are given by an estimated statistic which is corrupted by estimation errors due to noise or finite simple size effects. Thus, they are only ‘‘approximatively’’ jointly block diagonalizable as it will be discussed in the following.

\subsubsection{Approximate joint block-diagonalization}
It is known \cite[e.g.][]{nolte06,meinecke12} that the problem of the approximate joint block-diagonalization of a set of real-valued and antisymmetric matrices can be solved by transforming it in an ordinary approximate joint diagonalization problem, for which a number of popular and computationally appealing algorithms are available \cite[e.g.][]{cardoso96,afsari06,theis06,li07,tichavsky12}, if complex-valued matrices are allowed for the diagonalizing matrices. Indeed, a complex-valued matrix $\mathbf{W}_2$ exists which diagonalizes, in the ordinary sense, all the 2$\times$2 block diagonal matrices, i.e.
\begin{equation}
\left\{\begin{array}{c}
\mathbf{W}_2 \mathbf{W}_1 \operatorname{Re}\left\{\mathbf{B}_{[:|j|:]}\right\}\mathbf{W}_1^\dagger \mathbf{W}_2^\dagger = \mathbf{\Lambda}^R_j\\
\mathbf{W}_2 \mathbf{W}_1 \operatorname{Im}\left\{\mathbf{B}_{[:|j|:]}\right\}\mathbf{W}_1^\dagger \mathbf{W}_2^\dagger = \mathbf{\Lambda}^I_j
\end{array}\right.
\label{sim_diag}
\end{equation}
for $j=1...N$, where $\mathbf{\Lambda}^R_j$ and $\mathbf{\Lambda}^I_j$ are diagonal matrices (i.e., having non-zero entries only on the main diagonal), or as diagonal as possible for approximate solutions, and
\begin{equation}
\mathbf{W}_2=\frac{1}{2}\mathbb{I}_{Q\times Q}\otimes \left( 
\begin{array}{cc}
1&-\jmath\\
1&\jmath
\end{array}\right)
\end{equation}
with $\mathbb{I}_{Q\times Q}$ being the identity matrix of size $Q$ and $\otimes$ the Kronecker product. Therefore, our problem is equivalent to estimating $\mathbf{W}=\mathbf{W}_2\mathbf{W}_1$ by means of ordinary approximate joint diagonalization of the above set of matrices, since from $\mathbf{W}^{-1}=\mathbf{W}_1^{-1}\mathbf{W}_2^{-1}$ we observe that: (i) the columns of $\mathbf{W}^{-1}$ come in pair, i.e. if $\mathbf{w}$ is a column, then so it is $\mathbf{w}^*$; and (ii) the desired source topographies are contained in the real and imaginary part of the columns of $\mathbf{W}^{-1}$, i.e.,
\begin{eqnarray}
&&\mathbf{W}^{-1}= \bigr[\mathbf{a}_1+\jmath\mathbf{b}_1 \,,\,\mathbf{a}_1 - \jmath\mathbf{b}_1\,,\, \dots \,,\, \mathbf{a}_Q+\jmath\mathbf{b}_Q \,,\, \mathbf{a}_Q-\jmath\mathbf{b}_Q\bigr] \label{invW_orig}
\end{eqnarray}
Since the approximate joint diagonalization procedures return in output a diagonalizing matrix $\mathbf{W}$ which has at most the same size of the input matrices, it follows that we can extract at most as many underlying sources as the number of sensors, from which the second assumption of biPISA (i.e. the number of interacting sources is not greater than the number of EEG or MEG channels) necessarily follows.

As mentioned above, various algorithms could be used to address the problem of ordinary approximate joint diagonalization. In this work, we used the Cardoso and Souloumiac's algorithm \cite{cardoso96}, which is restricted to the case where the diagonalizing matrix (and therefore its inverse) is unitary. Note that we could include the unitarity constraint because of an inherent non-uniqueness of the solution to the joint diagonalization problem. Indeed, if $\left\{\mathbf{W},\mathbf{\Lambda}_j\right\}$ is a solution, then $\left\{ \mathbf{W}'= \mathbf{\Gamma}\mathbf{U}\mathbf{W} , \mathbf{\Lambda}'_j=\mathbf{\Gamma}\mathbf{U}\mathbf{\Lambda}_j\mathbf{U}^\dagger\mathbf{\Gamma}^{-1} \right\}$ is another admissible solution, with $\mathbf{U}$ being a unitary matrix and $\mathbf{\Gamma}$ being a diagonal matrix. 
This implies that what we get in practice is not the matrix $\mathbf{W}^{-1}$, but its orthogonal basis matrix $\mathbf{W}'^{-1}$.\\
The main consequence of the above choice is that, under unitarity constraint on the diagonalizing matrix, we are no more able to straightforwardly retrieve the exact topographies of interacting sources, but only the subspace they span, as explained in the following. We observe from $\mathbf{W}'^{-1} = \mathbf{W}_1^{-1}\mathbf{W}_2^{-1}\mathbf{U}^{\dagger}\mathbf{\Gamma}^{-1} $ (where we remind that $\mathbf{W}_1^{-1}=\mathbf{A}$ is the real-valued matrix of source topographies) that the columns of $\mathbf{W}'^{-1}$ are still a linear combination of the columns of $\mathbf{W}_1^{-1}$, with unknown complex-valued coefficients, i.e. the entries of $\mathbf{W}_2^{-1}\mathbf{U}^{\dagger}\mathbf{\Gamma}^{-1}$. If we then construct a real-valued matrix $\mathbf{X}$ by concatenating the real part and the imaginary part of $\mathbf{W}'^{-1}$, i.e. $\mathbf{X}=\left[ \operatorname{Re}\left\{\mathbf{W}'^{-1}\right\} \operatorname{Im}\left\{\mathbf{W}'^{-1}\right\} \right]$, we get that the first $M$ left-singular vectors of $\mathbf{X}$ span the same subspace of the columns of $\mathbf{W}_1^{-1}$, i.e. the source topographies. In practice, $M$ is unknown and, without a priori knowledge, it is set equal to $N$.

In order to retrieve the actually interacting source topographies from the above subspace, additional assumptions are required, as explained in section \ref{sourceretr_sec}.

\subsubsection{Dimensionality reduction prior to approximate joint block diagonalization\label{svddimred_sec}}
Before dealing with the issue of retrieving the interacting sources from the compound subspace they span, we describe a dimensionality reduction procedure which can be performed prior to the approximate joint block diagonalization. The aim of this procedure is to identify a smaller set of meaningful matrices to block diagonalize as an alternative to all the $j$th slices of the antisymmetric bispectral tensor $\mathbf{B_{[i|j|k]}}$, which allows for better performances of the diagonalization algorithm.\\
We first unfold the tensor $\mathbf{B_{[i|j|k]}}$ by reordering the element of each $j$-th slice in $N^2$-length vectors which form the column of an $N^2$$\times$$N$ matrix, namely $\mathbf{L}$. Then, a singular value decomposition (SVD) of $\mathbf{L}$ is computed, i.e., $\mathbf{L}=\mathbf{U\Sigma V^{\dagger}}$, and the left-singular vectors forming the columns of $\mathbf{U}$ (eventually multiplied by $\mathbf{\Sigma}$) are refolded in new square matrices, forming the $j$the slices of a novel tensor $\mathbf{\tilde{B}_{[i|j|k]}}$, which will be considered for diagonalization. It follows, indeed, from the properties of the SVD that the $j$th slices of $\mathbf{\tilde{B}_{[i|j|k]}}$ are weighted sums of the $j$th slices of $\mathbf{B_{[i|j|k]}}$, with weights given by the elements of the corresponding vector in $\mathbf{V}$. Thus, they are also diagonalized by $\mathbf{W}$. In addition, they are ordered according to descending singular values, then the first matrix has the maximal contribution to the norm of the original tensor $\mathbf{B_{[i|j|k]}}$, the second optimizes norm subject to being orthogonal to the first, and so on. Here, the orthogonality is meant with respect the scalar product of the singular vectors which generate the matrices. The main advantage of this procedure is that we can choose to diagonalize a smaller subset of the initial matrices, ignoring those matrices whose contribution to the bispectral tensor is negligible. In practice, the threshold is set by looking at the normalized version of individual matrices according to the normalization introduced in equation \ref{acb_norm}. Indeed, it also follows from multilinear properties of cumulant tensors that the elements of the new tensor $\mathbf{\tilde{B}_{[i|j|k]}}$ are the antisymmetric components of the cross-bispectrum between $x_i(f_1)$, $x_k(f_1+f_2)$ and a weighted sum of all the other channels at frequency $f_2$, with weights given in $\mathbf{V}$, from which the expression in equation \ref{acb_norm} can be evaluated.  

\subsubsection{Pairwise interacting sources retrieval and phase-delay estimation\label{sourceretr_sec}}
The problem of decomposing the obtained $M$-dimensional subspace into the contributions of the individual brain sources is now addressed with the minimum overlap component analysis (MOCA) \cite{marzetti08,nolte09}. The main idea underlying MOCA is to assume that the vector fields of the localized brain sources, i.e. after an inverse solver has been applied to source topographies, are (maximally) spatially separated. Here, MOCA is applied to the set of singular vectors identified by biPISA and the resulting topographies are subsequently interpreted as the topographies of the actually interacting sources, i.e. the columns of the matrix $\mathbf{A}$. We finally recognize among the separated source topographies those which form the interacting pairs by testing the arrangement that best block diagonalizes the above set of matrices and get the final form for the matrix $\mathbf{A}$.

Once the matrix $\mathbf{A}$ has been retrieved, an approximate estimate of the source level tensor $\boldsymbol{\mathcal{B}_{[m|n|p]}}$ can be obtained by inverting the two-step computation depicted in figure \ref{tens_mult}. This allows to retrieve the complex-valued coefficients $\alpha_q$ and $\beta_q$, $q=1...Q$, which provide further information on the interacting subsystems. For instance, we note that the magnitude of these coefficients measures the strength of the interaction. Then, we might define an index of interaction
\begin{equation}
\varepsilon_q=\frac{|\tilde{\alpha}_q|^2+|\tilde{\beta}_q|^2} {\sum_{q^\prime=1}^Q | \tilde{\alpha}_{q^\prime} |^2+| \tilde{\beta}_{q^\prime}|^2}
\end{equation}
which is the fraction of the interaction which is accounted for the $q$th subsystem. Furthermore, taken individually, the coefficients $\alpha_q$ and $\beta_q$ reveal how the interacting frequency components are distributed between the two sources. If, for instance, the sources $s_{1q}$ and $s_{2q}$ have distinct frequency content and, say, $s_{1q}$ contains only the component at frequency $f_1$ and $s_{2q}$ contains the remaining components at frequencies $f_2$ and $f_1+f_2$, then $\alpha_q$ would vanish, while $\beta_q$ would be non-vanishing and it would be equal to the conventional cross-bispectrum $\mathcal{B}_{122}$ between $s_{1q}$ and $s_{2q}$. On the contrary, if both sources contain all the three frequency components, then both $\alpha_q$ and $\beta_q$ would be non-vanishing, and they would reflect the relationship between different combinations of these components in the two sources. 

We will now elaborate more on the latter case by considering a special scenario in which the interaction within each subsystem consists in the coupling between two sources having an inherent cross-frequency coupling and being time delayed copies of each other, i.e.,
\begin{eqnarray}
s'_{2q}(t) = C_q\; s'_{1q}(t-\tau_q) \;\; \Rightarrow \;\; s_{2q}(f) =C\; s_{1q}(f)\; e^{- \imath 2 \pi f \tau_q} \label{time_delay}
\end{eqnarray} 
for $q=1...Q$, where $s_{1q}'(t)$ and $s_{2q}'(t)$ denote the activities in the time domain of the of sources belonging to the $q$th pair, $s_1q(f)$ and $s_2q(f)$ their respective Fourier transforms, $C_q$ is a real scale factor and $\tau_q$ is a non-zero time delay. Then, by analogy to what has been derived elsewhere \cite{chella14} in relation to an interaction model consisting of only one pair of interacting neuronal sources, rather than $Q$ pairs as in biPISA, the argument of the ratio between $\alpha_q$ and $\beta_q$ provides an estimation of the phase difference between components at frequency $f_2$ of the sources belonging to the $q$th pair, i.e.,
\begin{equation}
\Delta\phi_q(f_2,\tau_q) = \arg \left( \frac{\alpha_q(f_1,f_2)}{\beta_q(f_1,f_2)} \right)
\label{phase_ext}
\end{equation}
We emphasize that the above result strictly relies on the assumptions of our model. If these are not met and, for instance, one source lacks the component at frequency $f_2$, then $\alpha_q$ or $\beta_q$ would vanish, and any attempted estimation of $\Delta\phi_q$ would return unreliable and inconsistent results over repeated experiments in the same or different subjects.

Finally, we observe that the above theory has been derived for the decomposition of the antisymmetric bispectral tensor evaluated at a frequency pair alone. This meets many practical needs, as we always observed the nonlinear phenomena revealed by antisymmetric bispectral analysis to be very specific in frequency, i.e. showing narrow peaks in the 2d frequency plane. Nevertheless, the extension to a wide-band interaction analysis is fairly straightforward, i.e. by requiring the joint block-diagonalization of all the $j$th slices from multiple tensors estimated at different frequency pairs, although the increase of the computational efforts should be considered, i.e., for the joint diagonalization procedure.

\section{Results\label{results_sec}}
\subsection{Simulations\label{subsec:simul}}
The performances of biPISA were evaluated by using numerical experiments. Ten minute MEG recordings, sampled at 500 Hz, were generated by using a realistic standard head model \cite{fonov09,fonov11} and a 153-channel sensor array, whose geometry faithfully reproduced the sensor layout of the whole-head MEG system installed at the University of Chieti \cite{pizzella01,chella12}. Virtual head location with respect to the sensors was taken from a real ordinary experiment. For each simulation repetition, the activities of a set of neuronal sources were generated and channel recordings were numerically computed by solving the MEG forward problem. In particular, all sources were modeled as single current dipoles randomly located at the vertices of a regular 5 mm spaced grid covering the whole brain volume. A minimum distance of 1 cm between sources was also required. The leadfield matrix was computed with the FieldTrip software package \cite{oostenveld11} by using a realistically shaped single-shell volume conduction model \cite{nolte03}. The set of sources included both interacting sources and sources of noise, the latter aiming at mimicking the background brain activity, which are described below.

\begin{figure*}[!p] 
\centering
\squeezetable
\includegraphics[width=14.5cm]{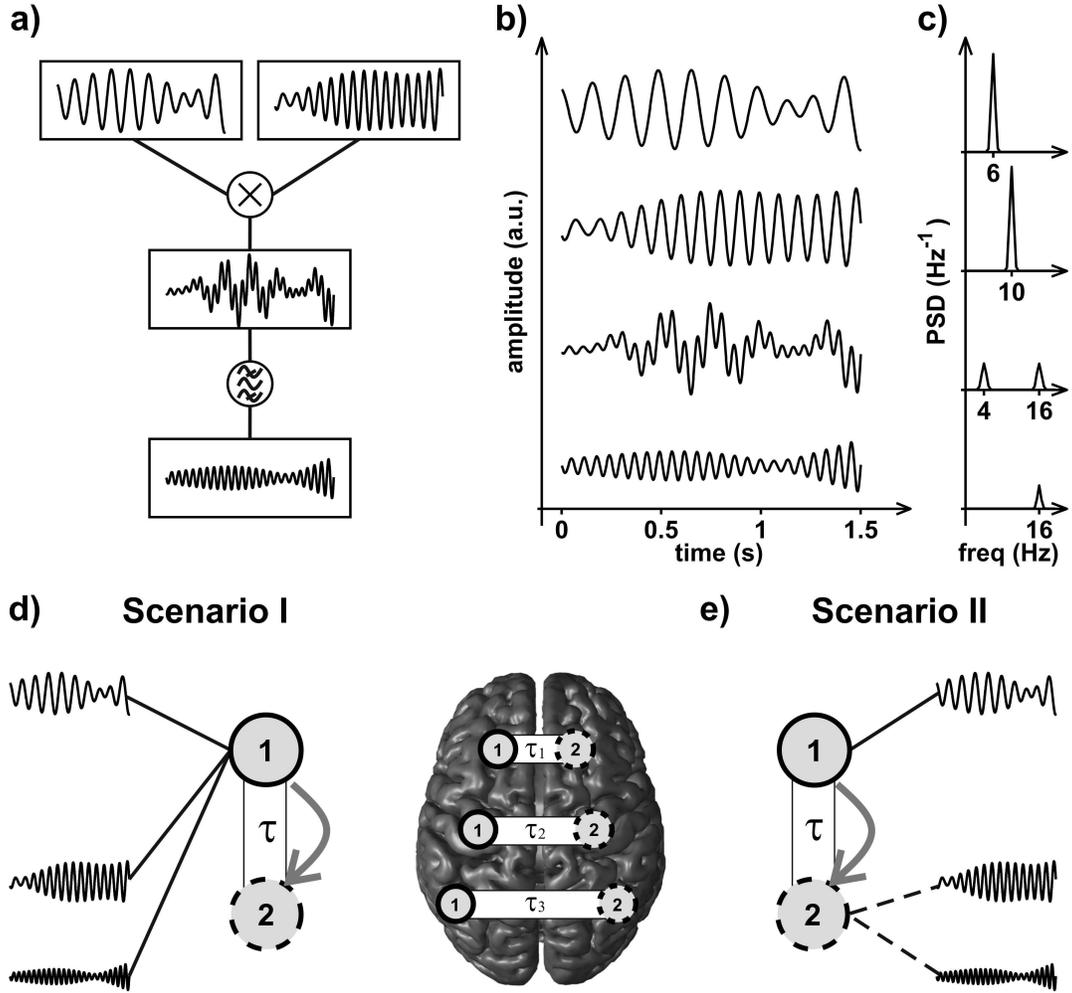}
\caption{\small{Simulated data generation. Panel a: two independent oscillators at 6 Hz and 10 Hz (top signals) were generated by band-pass filtering white Gaussian noise around the respective frequencies, with 1 Hz bandwidth. A 16 Hz oscillator (bottom signal) was generated through a multiplicative interaction, i.e. a time-point by time-point multiplication, between the oscillators at 6 Hz and 10 Hz, followed by band-pass filtering around 16 Hz with 1 Hz bandwidth. Short segments of the oscillator timecourses and the respective power spectral densities (PSD) are shown in panels b and c. The oscillators generated in this way resulted in quadratic phase coupling, and they were used for the construction of pairwise interacting source timecourses according to two different scenarios of interaction. In the first scenario (panel d), the timecourse of source 1 was obtained by summing the timecourses of all the three oscillators, and the timecourse of source 2 was set to a time delayed copy of the timecourse of source 1, with a time delay $\tau$. Up to 3 interacting source pairs were generated, with $\tau$ being 5 milliseconds for the first pair, 10 milliseconds for the second pair, and 15 milliseconds for the third pair. As a result, both sources contained the three components at frequencies 6 Hz, 10 Hz and 16 Hz. In the second scenario of interaction (panel e), the timecourse of source 1 was set to the timecourse of the oscillator at 6 Hz, whereas the timecourse of source 2 was set to the sum of the timecourses of the 10 Hz and 16 Hz oscillators. In order to account for a time delay between the two sources, in this scenario of interaction the 16 Hz oscillator was generated by multiplying the oscillator at 10 Hz by a time-delayed copy of the oscillator at 6 Hz, with the time delay $\tau$ being as in the previous case.}\label{fig:sim:illustration}}
\end{figure*}

The \textit{interacting sources} consisted of $Q$ interacting source pairs. In order to investigate different levels of interaction complexity, the number of pairs, $Q$, was also varied from 1 to 3. Each pair of interacting sources consisted of two dipoles exhibiting a cross-frequency phase interaction between components with frequencies $f_1=6$ Hz, $f_2=10$ Hz and $f_3=f_1+f_2=16$ Hz. Two possible scenarios of interaction were investigated, which are schematically depicted in figure \ref{fig:sim:illustration}.
The first scenario (scenario I) was formulated based on the time-delayed interaction model of equation \ref{time_delay}, and consisted in two sources having an inherent cross-frequency coupling (for which they will be referred to as `nonlinear sources' in the following) and being time delayed copies of each other. In particular, the time delay was introduced to mimic the actual delay in the information flow from the first to the second source due to the separation of sources in space and given the limited transmission speed \cite{ringo94,swadlow00}. This implies a phase difference between the components of the same frequency in the two sources, whose estimation using biPISA was one of the objective of this investigation. Thus, for each $q$th source pair, with $q=1...Q$, we first generated the timecourse of a nonlinear source, $s'_{1q}(t)$, and then we set the timecourse of a second source, $s'_{2q}(t)$, to a time-delayed copy of $s'_{1q}(t)$, i.e. $s'_{2q}(t) = s'_{1q}(t-\tau_q)$. The time delays were set to: (i) 5 milliseconds in case of only one source pair ($Q$=1); (ii) 5 and 10 milliseconds for the first and second pair, respectively, in case of two source pairs ($Q$=2); and (iii) 5, 10 and 15 milliseconds for the first, second and third pair, respectively, in case of three source pairs ($Q$=3). The timecourse of the nonlinear source was generated by summing the timecourses of two independent oscillators at 6 Hz and 10 Hz, i.e. obtained by band-pass filtering white Gaussian noise around 6 Hz and 10 Hz, respectively, and the timecourse of a 16 Hz oscillator which was phase-synchronous to the former, i.e. resulting from a multiplicative interaction process (namely, a time-point by time-point multiplication) between the other two oscillators, followed by filtering around 16 Hz. For the filtering at the above three frequencies, we used a Butterworth filter with 1 Hz bandwidth, performing filtering in both the forward and reverse directions to ensure zero phase distortion. In this scenario of interaction, each of the two sources belonging to the same pair contained the three frequency components at 6 Hz, 10 Hz and 16 Hz.\\
The second scenario (scenario II) of interaction consisted in a pure cross-frequency coupling between sources, that is, for each $q$th source pair, one source contained only the frequency component at 6 Hz, whereas the remaining frequency components at 10 Hz and 16 Hz were contained in the second source. This was obtained by setting the timecourse of the first source, $s'_{1q}(t)$, to the timecourse of the oscillator at 6 Hz and the timecourse of the second source, $s'_{2q}(t)$, to the sum of the timecourses of the 10 Hz and the 16 Hz oscillator. Unlike the previous scenario, and in order for the model to include a delay in the information flow from the first to the second source, in the present scenario the 16 Hz oscillator was generated by multiplying the oscillator at 10 Hz by a time-delayed copy of the oscillator at 6 Hz. This corresponds to the following relationship:
\begin{equation}
\tilde{s}'_{2q}(t) = s'_{2q} + h'(t) \ast \left[ s'_{1q}(t-\tau_q)\,s'_{2q}(t)\right]
\label{eq:scenarioII}
\end{equation}
where $s_{1q}'(t)$ and $s_{2q}'(t)$ denote the timecourses if the `uncoupled' source components at 6 Hz and 10 Hz, respectively, belonging to the $q$th pair ($q=1...Q$), $\tilde{s'}_{2q}(t)$ denotes the timecourse of the second source resulting from the coupling to the first source, $\tau_q$ is the time delay, $h'(t)$ is a transfer function for the filtering around 16 Hz, and $\ast$ denotes convolution operation. The values for $\tau_q$ were equal to the ones used in the previous scenario.

The \textit{sources of noise} consisted in 4 uncorrelated nonlinear sources exhibiting an inherent cross-frequency phase synchronization at the same frequencies as the interacting sources. The choice of this specific noise model was motivated by the fact that, as demonstrated elsewhere \cite{chella14}, for finite length data the presence of nonlinear noise, rather than other kinds of noise, e.g. Gaussian noise, still affects the estimation of the antisymmetric components of the cross-bispectrum and, thus, is more appropriate in our simulations.

For each simulation repetition, channel recordings were determined by varying independently: (i) the number of interacting source pairs, i.e. $Q$=1, 2 or 3; and (ii) the signal-to-noise ratio, i.e. SNR=$\infty$, 10, 2 or 1, in order to explore no-noise, low-noise, medium-noise and high-noise conditions. In particular, the SNR was calculated as the ratio between the mean variance across channels of the signal generated by interacting sources and the mean variance of the signal generated by nonlinear noisy sources. In order to simulate real experimental conditions, channel recordings were also contaminated by a low level of Gaussian noise. A total of 1000 simulation repetitions for each scenario of interaction was performed by randomizing dipole locations and orientations.

Bispectral analysis was firstly preceded by a dimension reduction stage using Principal Component Analysis (PCA) in order to lower the computational costs required to estimate the antisymmetric components of the cross-bispectrum between all channel triplets. The original set of 153 channel recordings was, thus, reduced to a smaller dataset including the first 30 principal components. The obtained signals were divided into 1 second non-overlapping segments. Within each segment, data were Hanning windowed, Fourier transformed and the antisymmetric bispectral tensor $\mathbf{B_{[i|j|k]}}$ was estimated at the frequency pair $(f_1,f_2)=(6\mbox{Hz},10\mbox{Hz})$, namely at frequency pair where interaction was simulated.

\begin{figure*}[!htb] 
\centering
\includegraphics[width=16.5cm]{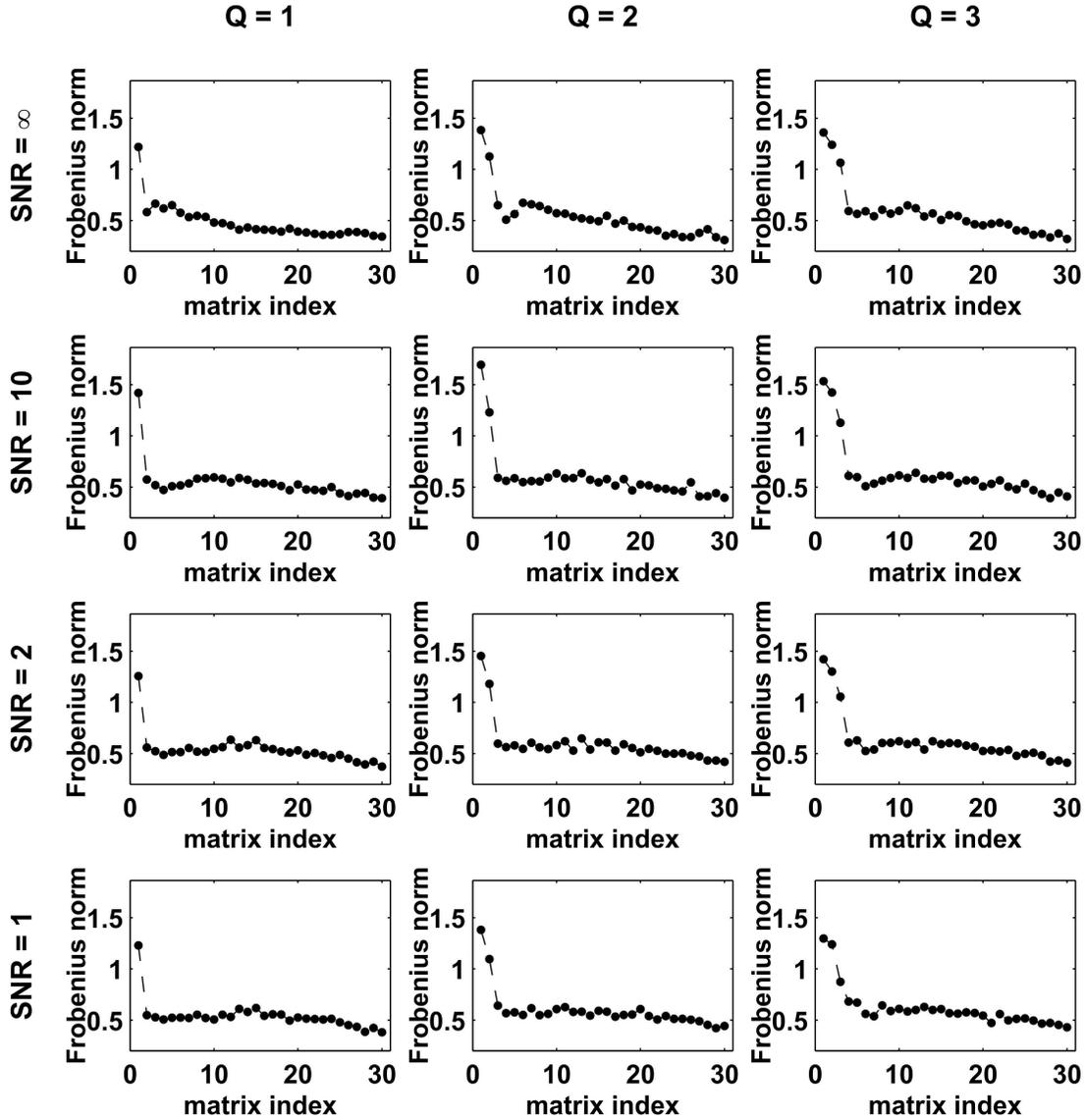}
\caption{\small{The plots show, for all combinations of pair numbers, i.e. $Q$, and SNRs, the Frobenius norms of the 30 normalized matrices (ordered in abscissa), i.e. whose entries have been normalized as in equation \ref{acb_norm}, returned by the SVD-based factorization described in the paper. Data are from one representative simulation repetition, corresponding to the scenario I of interaction between sources.}\label{fig:sim:svd}}
\end{figure*}

A second dimension reduction stage was performed by using the SVD-based factorization described in section \ref{svddimred_sec}, where the main aim was to identify a small set of matrices to diagonalize as an alternative to all of the $j$th slices of $\mathbf{B_{[i|j|k]}}$. An illustrative example of the results obtained at this stage of the analysis is given in figure \ref{fig:sim:svd}.
Here, for each of the matrices returned by the SVD-based factorization (ordered in abscissa) and for all combinations of pair numbers and SNRs, we show the Frobenius norms of the corresponding normalized matrices, i.e. whose entries have been normalized as in equation \ref{acb_norm}. In all cases, we observe a few values which can be clearly distinguished from the others. The corresponding matrices were, thus, interpreted as those containing the essential part of the interaction and considered for diagonalization. We also note that the number of these values always equates the number of simulated source pairs. Notably, this equivalence was observed for both of the two scenarios of interaction considered in simulations. This result is not trivial, but it could be easily tested to be true in the ideal case when, in equation \ref{gen_acb}, the coupling terms between sources belonging to different pairs completely vanish. On the basis of this result, in the following analysis, we focused on a number of sources, $M$, being twice as large as the number of selected matrices.

\begin{figure*}[!htb]
\centering
\includegraphics[width=16.5cm]{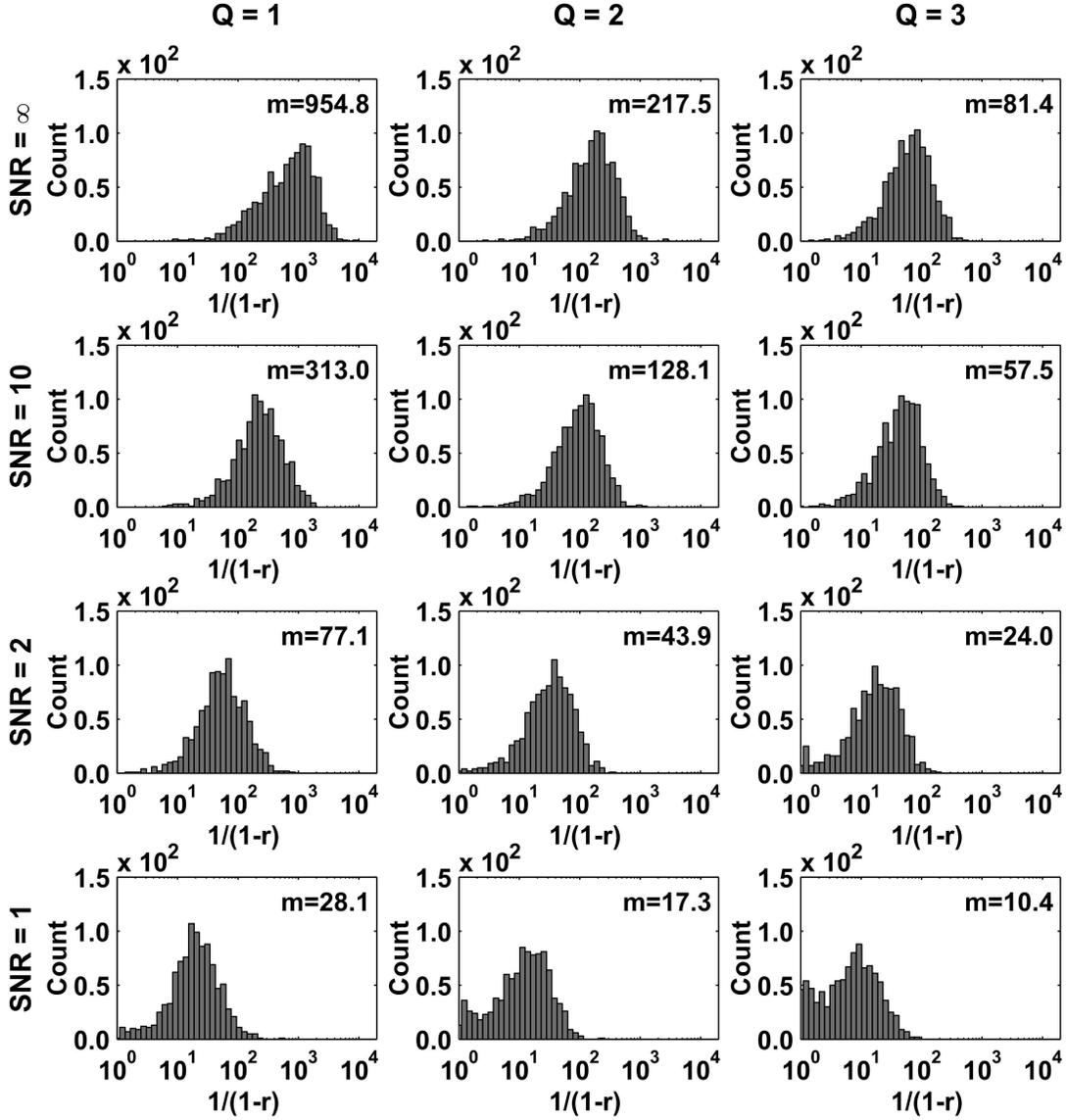}
\caption{\small{Scenario I of interaction. Histograms of $1/(1-r)$, with $r$ being the smallest of the $M$ canonical correlation coefficients between the true and estimated $M$-dimensional subspaces spanned by interacting source topographies, for all combinations of pair numbers, i.e. $Q$, and SNRs. \textit{m} denotes the mean value. Data from 1000 simulation repetitions.}\label{fig:sim:ccI}}
\end{figure*}

\begin{figure*}[!htb]
\centering
\includegraphics[width=16.5cm]{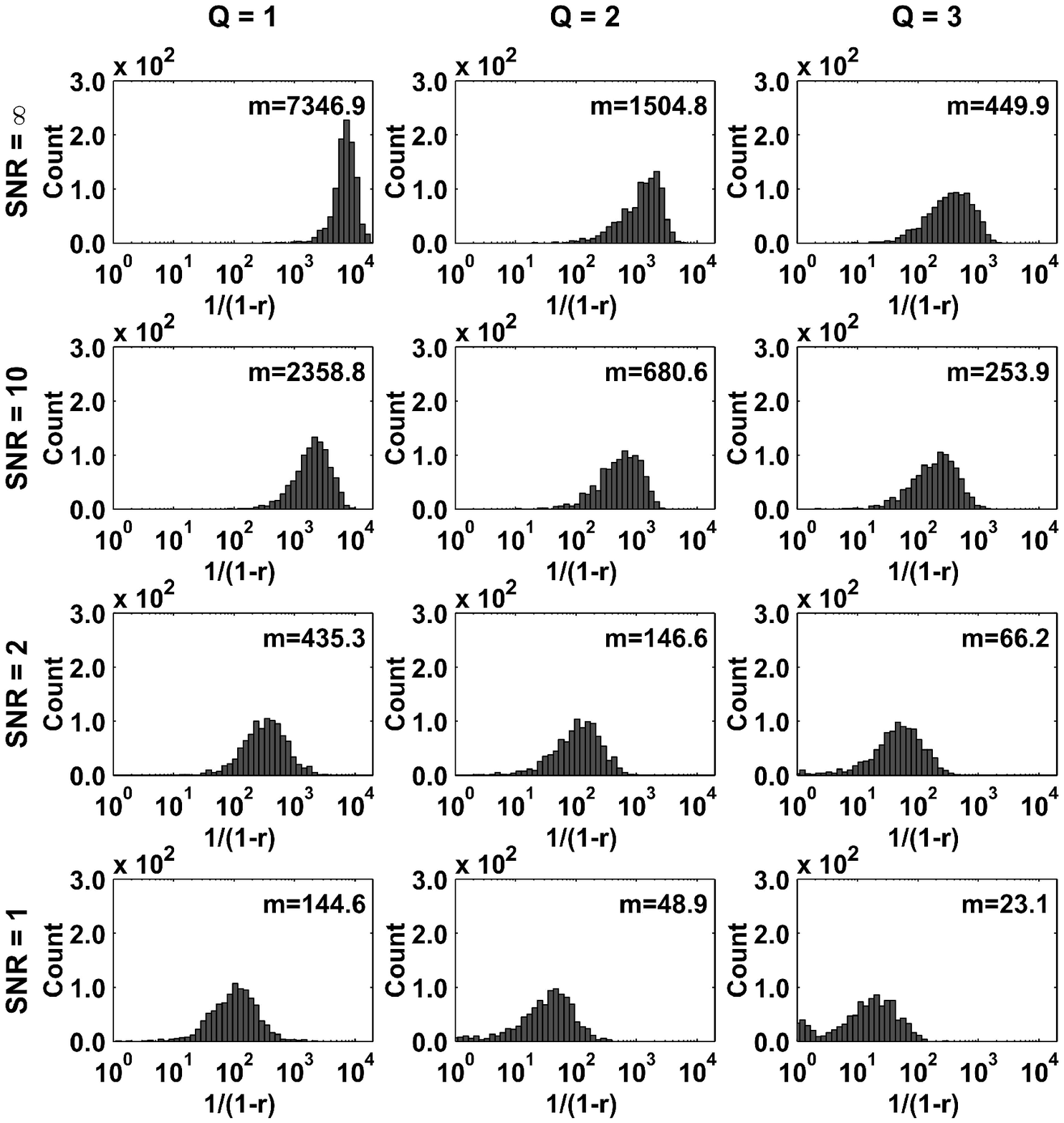}
\caption{\small{Scenario II of interaction. Histograms of $1/(1-r)$, with $r$ being the smallest of the $M$ canonical correlation coefficients between the true and estimated $M$-dimensional subspaces spanned by interacting source topographies, for all combinations of pair numbers, i.e. $Q$, and SNRs. \textit{m} denotes the mean value. Data from 1000 simulation repetitions.}\label{fig:sim:ccII}}
\end{figure*}

Next, the diagonalization of the real and imaginary part of the matrices resulting from the SVD-based factorization was performed to estimate the diagonalizing matrix $\mathbf{W}'$. The first $M$ columns of its inverse, $\mathbf{W}'^{-1}$, ordered according to descending diagonal elements, were then used to estimate the singular vectors spanning the subspace of the interacting source topographies. At this stage, i.e. before applying MOCA, we are not able to recognize the contribution of each individual source. We can nevertheless evaluate the performance of biPISA by looking at a measure of the similarity between the true (known in simulation) an the estimated subspace. In particular, the similarity was measured as the smallest of the $M$ canonical correlation coefficients between the two $M$-dimensional subspaces. The results are summarized in figure \ref{fig:sim:ccI} and figure \ref{fig:sim:ccII} for the scenario I and for the scenario II of interaction, respectively, where we show the histograms of $1/(1-r)$, with $r$ being the smallest of the $M$ canonical correlation coefficients, for all combinations of pair numbers and SNRs.

Overall, we observe that biPISA provides reliable estimates of the subspace spanned by the interacting source topographies in both of the simulated scenarios of interactions. Indeed, the correlation coefficient $r$ (respectively $1/(1-r)$) is, on average, greater than 0.9 (respectively 10) for all the investigated conditions. The performances are slightly affected by the complexity of interaction, here measured by the number of simulated source pairs, while the main downgrade is due to the increasing level of noise. The latter effect was explained as evidence that, for finite length data, the symmetric components of cross-bispectra arising from noisy sources may be not suppressed completely, thus affecting the result of any following data analysis, i.e. the joint diagonalization of bispectral matrices in this work. In the contrast between the performances of biPISA in the two simulated scenarios of interaction, we note that, for fixed values of Q and SNR, the correlation coefficients are systematically larger in the scenario II than in the scenario I of interaction. Indeed, the values of $1/(1-r)$ obtained in the scenario II (figure \ref{fig:sim:ccII}) are, on average, from $2.2$ ($Q$=3, SNR=1) up to $7.7$ ($Q$=1, SNR=$\infty$) times larger than the respective values obtained in the scenario I (figure \ref{fig:sim:ccI}). This was explained by the fact that in the latter case there was a cross-frequency coupling within the sources belonging to the same pair, and thus the interacting sources themselves introduced noise components due to their inherent cross-frequency coupling.

\begin{figure*}[!htb]
\centering
\includegraphics[width=8cm]{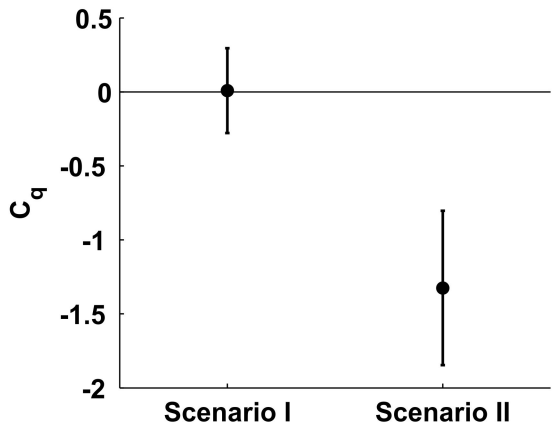}
\caption{\small{Contrast $C_q$ between the magnitudes of the coefficients $\alpha_q$ and $\beta_q$ obtained in the two simulated scenarios of interaction. The black dots denote the mean value; the error bars denote the standard deviations.}\label{fig:sim:ab}}
\end{figure*}

In order to test on the synthetic data the property of the proposed approach of being sensitive to the distribution of the frequency components between the interacting sources, we considered the coefficients $\alpha_q$ and $\beta_q$ estimated in simulations. Indeed, as was argued in section \ref{sourceretr_sec}, $\alpha_q$ should vanish in the second scenario of interaction. We therefore looked at the contrast between the magnitudes of the coefficients $\alpha_q$ and $\beta_q$ obtained in the two different scenarios of interaction. Specifically, for each $q$th interacting pair, the contrast $C_q$ was evaluated as the logarithm of the ratio between the magnitudes of $\alpha_q$ and $\beta_q$, i.e.
\begin{equation}
C_q = \log_{10}\frac{|\alpha_q|}{|\beta_q|}
\end{equation}
The desired analysis was performed by collecting the results obtained by varying the number of actual interacting subsystems ($Q$ = 1,2,3) and the level of nonlinear noise corrupting the signals (SNR = $\infty$, 10, 2, 1), while we only distinguished between the two simulated scenarios of interaction. The results are summarized in figure \ref{fig:sim:ab}. In this plot, the black dots denote the mean values of $C_q$, and the error bars denote the respective standard deviations. 
We observe that, in the scenario I, the magnitudes of $\alpha_q$ and $\beta_q$ are, on average, comparable ($C_q$= 0.00$\pm$0.28; mean$\pm$st.dev.). This particular result is due to the fact that, in or simulations, the contribution of the two interacting sources to signals was rather balanced, i.e. the sources being exact copies of each other, and having random locations. On the contrary, in the scenario II, the magnitude of $\alpha_q$ is, on average, more than 10 times smaller than the magnitude of $\beta_q$ ($C_q$= -1.32$\pm$0.52; mean$\pm$st.dev.), which is in line with our hypothesis. Lower values for the magnitude of $\alpha_q$ were not obtained, conceivably due to a bias of the fit.

\begin{figure*}[!p] 
\centering
\includegraphics[height=16.5cm,angle=270]{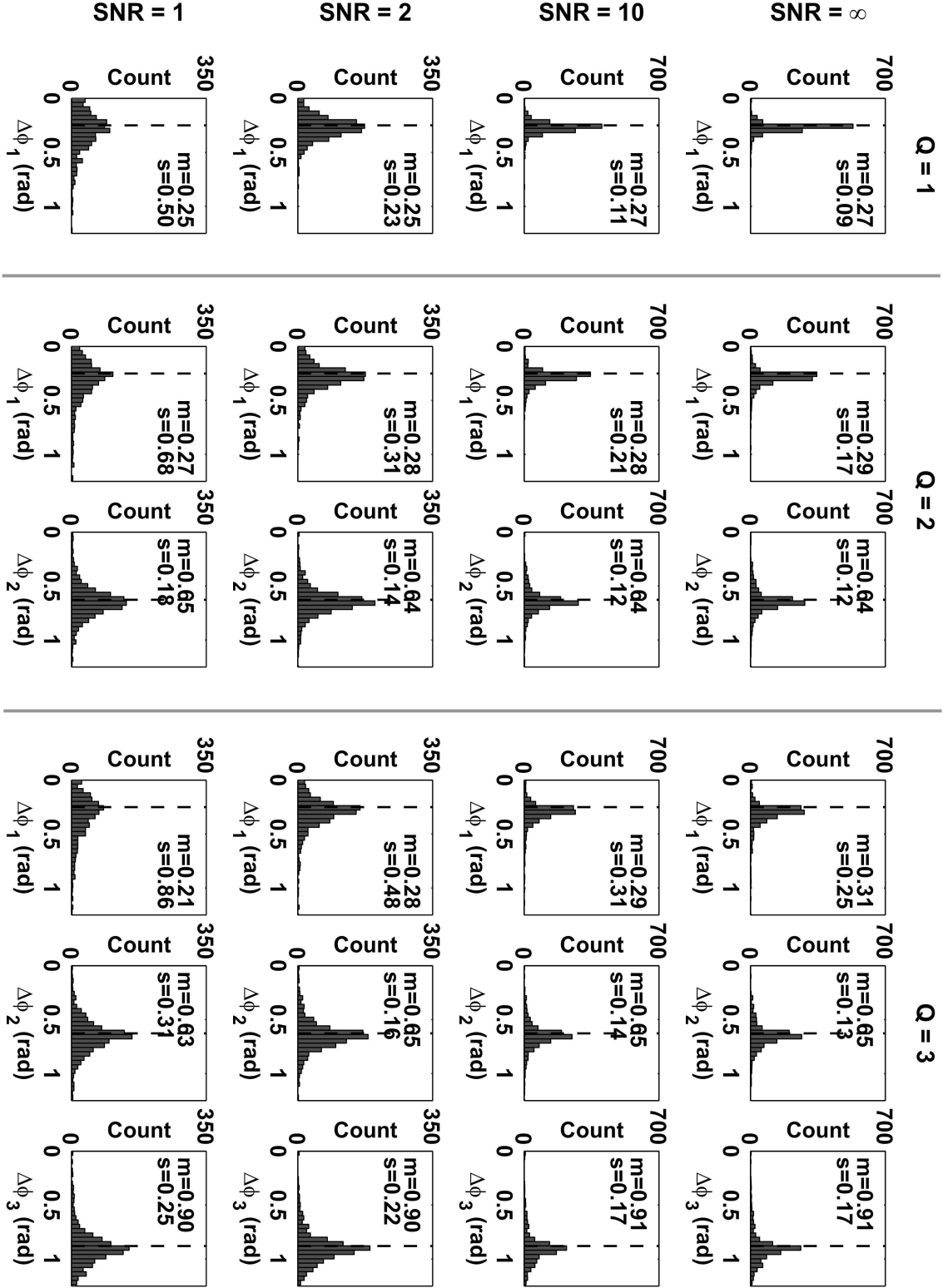}
\caption{\small{Scenario I of interaction. Histograms of the estimated phase differences between interacting sources, for all combinations of pair numbers, i.e. $Q$, and SNRs. The vertical dashed line denotes the true values for phase differences, i.e. $\Delta\phi_1=0.314$, $\Delta\phi_2=0.628$ and $\Delta\phi_3=0.942$. \textit{m} and \textit{s} denote the mean value and the standard deviation, respectively. Data from 1000 simulation repetitions.}\label{fig:sim:deltaphi}}
\end{figure*}

We finally evaluated the ability of biPISA in extracting reliable information about pairwise interaction dynamics, namely, after MOCA has been applied to disentangle individual source topographies, and after the interacting source pairs have been clustered by testing the arrangement which best block diagonalizes the above set of matrices. In particular, in relation to the data for the scenario I of interaction, we looked at the performances of the presented method in estimating the phase difference between the interacting sources. We then retrieved the antisymmetric bispectral tensor at source level and used the corresponding complex-valued terms $\alpha_q$ and $\beta_q$, with $q$ running over $1,...,Q=M/2$, to estimate the phase difference $\Delta\phi_q$, according to the expression in equation \ref{phase_ext}. The results are shown in figure \ref{fig:sim:deltaphi},
where the histograms of the values obtained in all simulation repetitions are plotted along with true simulated value, i.e. $\Delta\phi_1=0.314$ for the first pair, $\Delta\phi_2=0.628$ for the second pair, and $\Delta\phi_3=0.942$ for the third pair. We observe that, similarly to correlation coefficients, the results of phase estimation are moderately affected by the number of simulated pairs, while the main performance downgrade depends on the noise level, which is a fairly obvious conclusion, being this result dependent on the goodness of the source subspace retrieval.

\subsection{Application to real MEG data\label{subsec:realdata}}
As an example of application to real data, we applied biPISA to the analysis of MEG data recorded in one healthy adult subject (female; 22 years old; right handed) during 10 minutes of eyes-open resting state, while the subject was instructed to maintain fixation on a visual crosshair. MEG was recorded using the 165-channel MEG system installed at the University of Chieti \cite{pizzella01,chella12}. This system includes 153 dc-SQUID integrated magnetometers arranged on a helmet covering the whole head plus 12 reference channels. Signals were sampled at 1025 Hz. The position of the subject's head with respect to the sensors was determined by five coils placed on the scalp recorded before and after MEG recording. The coil positions were measured by a 3D digitizer (Polhemus, Colchester, VT, USA) together with anatomical landmarks (left and right preauricular and nasion) defining a head coordinate system. High resolution whole-head anatomical images were acquired using a 3-T Philips Achieva MRI scanner (Philips Medical Systems, Best, The Netherlands) via a 3D fast field echo T1-weighted sequence (MP-RAGE; voxel size 1 mm isotropic, TR = 8.1 ms, echo time TE = 3.7 ms; flip angle 8$^\circ$, and SENSE factor 2). The coregistration of the MEG sensor with the MRI volume was performed by aligning the anatomical landmarks in the two modalities.

After downsampling at 341 Hz and band-pass filtering at 1-80 Hz, data were preprocessed using an Independent Components Analysis (ICA), with the twofold purpose of removing the artifactual components, and reducing the dimensionality of the data. We found 25 ICs, which were visually inspected and classified as components of brain origin (18 out of 25) or artifactual components (7 out of 25). Typically, ICA based pipelines rely on the subtraction of artifactual ICs from MEG recordings. An alternative strategy is that of reconstructing MEG signals by recombining the ICs of brain origin \cite{mantini11,marzetti13}. By following the latter approach, we retained the ICs classified as brain components, which were given in input to bispectral analysis. Note that the maximum number of interacting sources that we can identify by using biPISA is now reduced to the number of retained ICs. 

\begin{figure}[!htbp] 
\centering
\includegraphics[width=8cm]{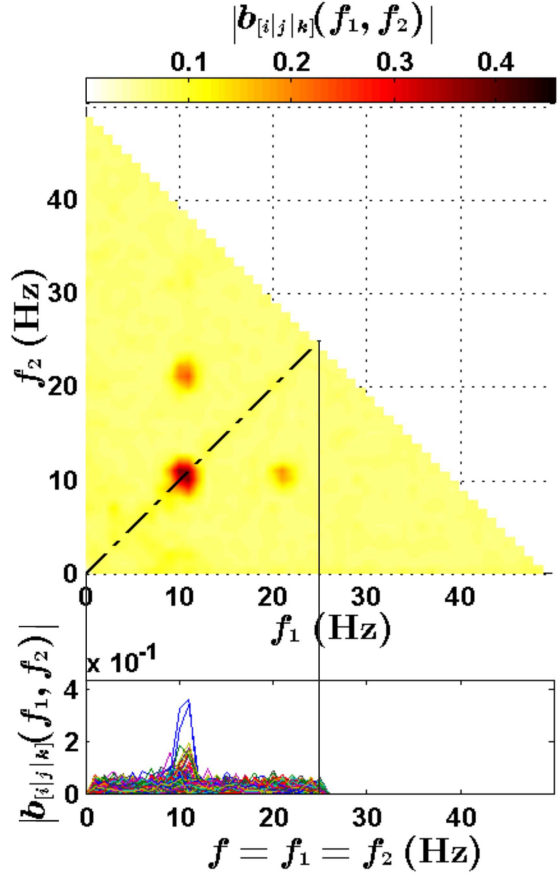}
\caption{\small{(Color online) On the top, the magnitude of the antisymmetric component of bicoherence, $\left|b_{[i|j|k]}(f_1,f_2)\right|$, is shown as function of frequencies $f_1$ and $f_2$. On the bottom, a detailed view of $\left|b_{[i|j|k]}(f_1,f_2)\right|$ for the diagonal axis $f_1=f_2$.}\label{fig:rdata:bic}}
\end{figure}

Bispectral analysis was performed by dividing signals into 1 second non-overlapping segments containing continuous data. Within each segment, data were Hanning windowed, Fourier transformed and the antisymmetric component of the cross-bispectrum (and bicoherence) was estimated for frequency pairs $(f_1,f_2)$ up to $f_1+f_2=50$ Hz. The resulting frequency resolution was 1 Hz on both the $f_1$ and $f_2$ axes. Figure \ref{fig:rdata:bic} shows the magnitude of the antisymmetric component of bicoherence, $\left|b_{[i|j|k]}(f_1,f_2)\right|$, as function of frequencies. We recall that $N^3$ estimates are obtained for each frequency pair, i.e. corresponding to all possible triplets formed by $N$ signals. However, only the maximum over these $N^3$ estimates can be appreciated from this plot. In order to gain more insight into the data, the values on the diagonal axis $f_2=f_1$, i.e. where the main interaction was found, are shown separately in the bottom part of the figure, i.e. where the reader can appreciate the values obtained for each channel triplet.
We observe a prominent peak at frequency pair (11Hz,11Hz), which reflects an interaction between frequency components at $f_1 = f_2=11 \text{ Hz}$ and $f_3=22 \text{ Hz}$. The antisymmetric bispectral tensor estimated at this frequency pair was then selected for analysis by using biPISA. 

\begin{figure}[!htbp] 
\centering
\includegraphics[width=8cm]{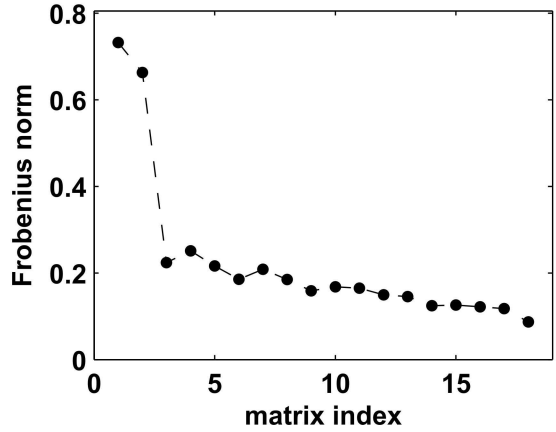}
\caption{\small{Frobenius norm of the normalized matrices (ordered in abscissa) returned by the SVD-based factorization of tensor slices for the antisymmetric component of the cross-bispectrum at $(f_1,f_2)$=(11Hz,11Hz).}\label{fig:rdata:svd}}
\end{figure}

\begin{figure}[!htbp] 
\centering
\includegraphics[width=8cm]{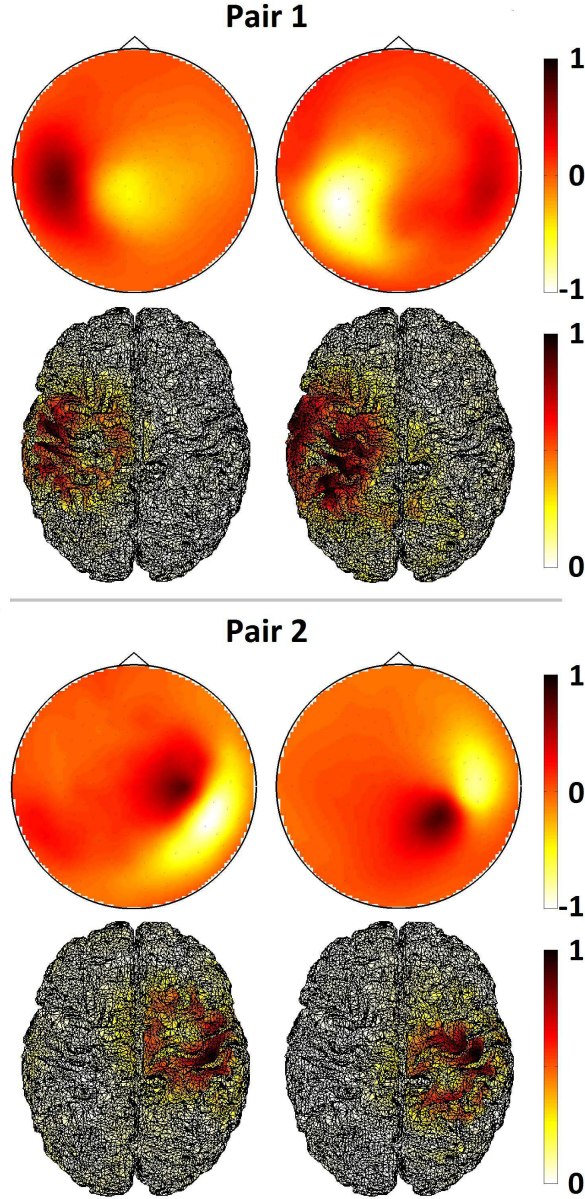}
\caption{\small{(Color online) Topographies (first and third lines) and the corresponding reconstruction of cortical activity (second and fourth lines) for pairwise interacting sources of brain mu (11 Hz) and beta (22 Hz) rhythms. The maps have been scaled between -1 and 1 for topographies, and between 0 and 1 for cortical activity. The maps use arbitrary units.}\label{fig:rdata:loc}}
\end{figure}

The SVD-based factorization of tensor slices revealed the existence of two components (see figure \ref{fig:rdata:svd}), which correspond to two pairwise interacting subsystems, and which clearly contain most of the observed interaction.      
The interacting sources were subsequently identified by simultaneous diagonalization, and finally separated by MOCA. The estimated source topographies are shown in figure \ref{fig:rdata:loc} (first and third lines). The respective source reconstructions, i.e. obtained by using a cortically constrained minimum norm estimate \cite{hamalainen93}, are shown below the topographies (second and fourth lines). 
Our findings clearly indicate an interaction between pairs of sources of central mu (11 Hz) and beta (22 Hz) rhythms, localizing in the proximity of the left and right central sulci. The interaction within the source pair localizing in the left hemisphere was stronger ($\varepsilon_1 =0.78$) than the interaction within the source pair localizing in the right hemisphere ($\varepsilon_2 =0.22$). The estimated phase differences were $\Delta\phi_1 = 0.73$ for the former source pair, and $\Delta\phi_2 = 0.80$ for the latter source pair.

\section{Discussion\label{discuss_subsec}}
We here proposed a method which allows to identify cross-frequency phase synchronous brain sources by decomposing the antisymmetric components of the cross-bispectra between EEG or MEG recordings. This approach, which we called biPISA, relies on the key assumption that the interactions between brain sources are pairwise. This is a clear simplification of the actual interaction dynamics, but it allows for a unique decomposition of the data in terms of brain sources and is able to capture the most relevant aspect of the interaction observed in EEG or MEG data. If this assumption is not met, the search for an approximate solution, i.e. by the approximate joint diagonalization algorithm, yields by construction the dominant part of the interaction, but the off diagonal terms contain additional information which will be neglected. We also assumed that the number of interacting sources is not grater than the number of recording channels. If this is violated, we would still observe interactions, but the decomposition in independent subsystems would be incomplete.

A key step of the analysis pipeline is the approximate joint diagonalization of the set of real-valued and antisymmetric matrices which come from the slices of the antisymmietric bispectral tensor at a certain frequency pair. To address this issue, we applied a well known algorithm for unitary approximate joint diagonalization \cite{cardoso96}, which allows to identify the subspace spanned by interacting source topographies, but the separation of individual sources, as well as the identification of interacting pairs, is not possible without first introducing additional (spatial) assumptions on the sources, i.e. the MOCA constraint. Of course, the above diagonalization algorithm could be replaced by others. For instance, the original PISA method \cite{nolte06} exploits a non-unitary diagonalization procedure, i.e. a generalization of the DOMUNG algorithm \cite{yeredor04} to the complex domain, to jointly diagonalize the set of antisymmetric cross-spectral matrices in a given frequency range. A non-unitary symplectic optimization algorithm was later proposed by \citet{meinecke12} to address the same issue. Both algorithms relax the unitarity assumption on the diagonalizing matrix, leading to a number of advantages over unitary transformations \cite{meinecke12}. Most notably, for both of these approaches, the joint diagonalization results in the separation of different pairs (but not in the separation of the two sources within each pair) if a wide-band analysis of the the data is available, while such a separation cannot be done for a single frequency alone. Within each pair, the separation of individual sources is only possible if additional constraints are introduced. Although the non-unitary methods could be conceptually more advantageous as they allow to straightforwardly separate the interacting subsystems, in this study, we used a naive unitary diagonalization procedure because of its computational efficiency and ease of implementation, leaving the interacting pair retrieval to a dedicated step performed after the demixing by MOCA. Obviously, in practical applications, this would be an advantage if the MOCA assumptions are met, while it would be a limitation in case of any actual overlap between sources belonging to different pairs.

It is worthwhile to address here the relationship between biPISA and the methods for Independent Component Analysis (ICA) which have, albeit with a different objective, some structural similarities. The ICA model assumes that the observed data result from an instantaneous mixing of statistically mutually independent sources \cite{comon94,hyvarinen00}. The methods for ICA aim at finding the independent sources by exploiting different properties of data derived from this assumption. A widely used class of ICA methods \cite[for a review see, e.g.,][]{comon10} exploits the fact that the cross-cumulants between independent sources are theoretically diagonal, and thus the sources can be found by the joint diagonalization of cross-cumulants between sensor data, i.e. cross-correlation matrices (e.g., the TDSEP \cite{ziehe98} algorithm), higher-order cumulant tensors, or even tensor-slices (e.g., JADE \cite{cardoso93}, or STOTD \cite{delathauwer01} algorithms). These methods assume that the cross-cumulants between sensor data are symmetric under permutation of their indices, which results from these quantities being a multilinear transformation of the diagonal cross-cumulants between sources. For instance, the TDSEP algorithm uses symmetrized versions of correlation matrices. Now, the main point here is that biPISA, whose aim is to find interacting sources - and not independent sources -, turns the above argument around. Specifically, in biPISA we assume that sources are interacting, which implies that the cumulants (and cumulant spectra) are not symmetric. Thus, in order to study interactions, we specifically focus on that part which deviates from symmetry. Indeed, we look at the antisymmetric component of the cross-bispectrum tensor, while we reject the symmetric component. We emphasize that in biPISA the independence assumption is only invoked for pairs, i.e. leading to a theoretically block-diagonal (but still not symmetric) model for source cumulants. This is the distinctive feature of the PISA approach, which we are now generalizing to include third-order spectra.

The effectiveness of the proposed approach was first tested in simulations. In particular, we investigated two possible scenarios of interaction between sources: (i) a time delayed interaction between nonlinear sources; and (ii) a pure cross-frequency interaction between sources. For each simulated scenario, we evaluated the ability of biPISA in identifying the interacting source pairs for different conditions constructed by varying the level of nonlinear noise corrupting the signals and the number of actual interacting subsystems. In addition and solely  for the second scenario of interaction, we evaluated the effectiveness in extracting information about the phase relationships between interacting sources. The results obtained in simulations showed that biPISA provides reliable results for all the investigated circumstances. We also observed that the performances worsen due mainly to the increasing level of nonlinear noise corrupting the signals, rather than to the increasing number of interacting subsystems. This is essentially due to the fact that, in the computation of the antisymmetric components of the cross-bispectra, the contribution of nonlinear noisy sources is suppressed in a statistical sense and, if too large, it may be not completely removed, thus affecting the result of the joint diagonalization procedure.

The analysis pipeline was applied to real MEG data recorded during eyes-open resting state. Our method was able to identify two pairs of brain sources exhibiting a cross-frequency phase synchronization between mu and beta rhythms, and localizing in the proximity of the left and right central sulci. This result is consistent with the findings of previous studies. Indeed, mu and beta rhythms have been repeatedly associated with the activity of the sensorimotor cortex \citep{salmelin94a,salmelin94b,pfurtscheller99}, and a functional relationship between these rhythms has also been strongly suggested \citep{ritter09}. Interestingly, mu and beta rhythms have been reported to originate in different areas of the sensorimotor cortex, i.e. the former in the post-Rolandic somatosensory area, and the latter in the pre-Rolandic motor area \citep{salmelin95,pfurtscheller99}. In this regard, it is important to note that our findings do not give specific information on the spatial segregation of these two rhythms. The issue of differentiating the spatial origin of mu and beta rhythms will be not addressed here, as a detailed physiological analysis of the observed phenomenon is beyond the scope of this paper. The possibility of separating the interacting sources on the basis of different criterion, e.g. disjoint supports in the frequency domain, will be addressed in future works.

\newpage

\section{Conclusions\label{conclusion_sec}}
In this paper, we presented a method, namely biPISA, which allows to identify the subsystems of larger system of interacting brain sources from the analysis of multichannel data, when the channel recordings are a linear, instantaneous and unknown superposition of sources activities, such as in EEG and MEG. In a broader perspective, the proposed method has potential applications in other areas for the analysis of multivariate data from complex systems, when only superimposed signals are available. In particular, superposition effects are dominant where only surface measurements are accessible, while the interesting dynamical processes, i.e. the sources, are hidden underneath, e.g., in acoustic, seismology, geophysics, astronomic or medical imaging.

The proposed approach is sensitive to nonlinear (i.e. cross-frequency) interactions between sources, which involve the synchronization between the phases of source oscillations at different frequencies. More specifically, the method relies on the estimation of the antisymmetric components of third-order statistical moments, i.e. cross-bispectra, between signals, with subsequent joint diagonalization of matrices constructed from these quantities. We emphasize that biPISA allows to reliably extract meaningful cross-frequency interaction while ignoring all spurious effects since, as opposed to conventional bispectral measures, the antisymmetric components of cross-bispectra cannot be generated from a superposition of non-interacting brain sources or other nonlinearities in the data, e.g. noise, but solely reflect the existence of genuine interactions. This method represents an extension to the analysis of cross-frequency brain interaction of a previous method, namely PISA (Pairwise Interacting Source Analysis) \cite{nolte06}, which was originally developed to investigate linear (i.e., frequency specific) brain interactions, i.e. involving phase coupling between oscillations at the same frequency, by decomposing the imaginary part of cross-spectral matrices. 

Simulated and real data analysis performed in this work revealed interesting features of brain interaction dynamic that may be captured by using the proposed approach. Taken altogether, our results demonstrate that biPISA can efficiently and effectively characterize cross-frequency couplings in brain networks by using noninvasive EEG or MEG measurements.

In conclusion, we believe that the proposed method might provide a new tool for gaining more insight into brain interaction dynamic and investigating the role of phase synchronization in the mechanisms of neuronal communications.


\appendix

\setcounter{section}{0}

\section{\label{AppA}}
Third order statistical moments in frequency domain can be written in the most general 
form as 
\begin{equation}
g(f_1,f_2,f_3)=\left< x(f_1)y(f_2)z^*(f_3)\right> \label{moments1}
\end{equation}
with $x$, $y$ and $z$ being the Fourier transforms of three signals in the time domain (here and in the following, lower case letters, e.g. $x$, will be used to denote signals in the frequency domain, while the corresponding signals in the time domain will be denoted in the same way with primed symbols, e.g. $x'$ ). The third signal, $z(f)$, was complex conjugated for convenience as becomes apparent below. With $z^*(f)=z(-f)$ this corresponds to the preferred definition of the sign of the frequency. 

For simplicity, we assume an odd number of time points and express the Fourier transformed signals by the original time series as
\begin{equation}
x(f_1)=\sum_{t_1=-N}^{N}x'(t)\exp\left(-\frac{2\pi i f_1t_1}{2N+1}\right)
\end{equation}
and analogously for $y$ and $z$.  Inserting this into (\ref{moments1}) we get
\begin{eqnarray}
g(f_1,f_2,f_3)=\sum_{t_1, t_2, t_3}\biggl[ \left<x'(t_1)y'(t_2)z'(t_3)\right> \nonumber \\
\qquad \qquad \cdot \exp\left(-\frac{2\pi i \left(f_1t_1+f_2t_2-f_3t_3\right)}{2N+1} \right) \biggr]
\label{moments2}
\end{eqnarray}
The crucial assumption is now that 
\begin{equation}
\left<x'(t_1)y'(t_2)z'(t_3)\right>=h'(t_1-t_2,t_1-t_3) 
\end{equation}
i.e., that this expectation value only depends on time differences and not on absolute time. This is the case 
for stationary processes, but,  in order to actually 
observe dependence on absolute time, also for non-stationary processes 
the clock defining absolute time has to 
be known and the analysis must be done relative to this clock. This is possible in an event-related experimental design, 
but for spontaneous brain activity and also in a task-related 
experimental design such a clock is not given and one cannot observe dependence on absolute time even if the process is truly 
non-stationary with respect to some hidden process. Also note that the above expectation value refers to the hypothetical situation   
 that the entire measurement can be repeated infinitely many times. In practice, this ensemble average is replaced by a time average 
over segments (with a much courser frequency resolution). Absolute time dependence can only be observed if the time relative to the beginning 
of each segment has a physical meaning like the time of a trigger.      

We now switch time coordinates and define $\tau_1=t_1$, $\tau_2=t_1-t_2$, and $\tau_3=t_1-t_3$ leading to
\begin{eqnarray}
f_1t_1+f_2t_2-f_3t_3 & = & (f_1+f_2-f_3)\tau_1 \nonumber \\
& & \qquad \qquad - f_2\tau_2+f_3\tau_3
\end{eqnarray}
Assuming that the time series is sufficiently long that we can ignore all edge-effects due to finite length of the data, we can rearrange the sums in (\ref{moments2}) to 
\begin{eqnarray}
g(f_1,f_2,f_3)&=\sum_{\tau_2,\tau_3}
h'(\tau_2,\tau_3)\exp\left(\frac{2\pi i \left(f_2\tau_2-f_3\tau_3\right)}{2N+1}\right) \nonumber \\
&\cdot \sum_{\tau_1=-N}^N \exp\left(\frac{2\pi i \tau_1\left(f_3-f_1-f_2\right)}{2N+1} \right)
\label{moments3}
\end{eqnarray}
The important point now is that 
\begin{eqnarray}
\sum_{\tau_1=-N}^N \exp\left(\frac{2\pi i \tau_1\left(f_3-f_1-f_2\right)}{2N+1} \right)= (2N+1)\delta_{0,f_3-f_1-f_2}
\end{eqnarray}
where $\delta$ denotes the Kronecker-delta function. Specifically, $g(f_1,f_2,f_3)$ can only be non-vanishing if $f_1+f_2=f_3$, which is what we wanted to show. 

\section{\label{AppB}}
The aim of this appendix is to demonstrate that if we define bicoherence as the cross-bispectrum divided by the symmetric part of \citet{shahbazi14} normalization factor, namely
\begin{equation}
b_{ijk}(f_1,f_2) =\frac{B_{ijk}(f_1,f_2)}{N_{(i|j|k)}(f_1,f_2)}
\label{bicdef:app}
\end{equation}
with
\begin{equation}
N_{(i|j|k)}(f_1,f_2) = N_{ijk}(f_1,f_2) + N_{kji}(f_1,f_2)
\end{equation}
and $N_{ijk}(f_1,f_2)$ being defined in equation \ref{uninorm}, then the antisymmetric component of bicoherence 
\begin{equation}
b_{[i|j|k]}(f_1,f_2) = b_{ijk}(f_1,f_2) - b_{kji}(f_1,f_2)
\label{abicdef:app}
\end{equation}
is a normalized version of the antisymmetric component of the cross-bispectrum, i.e.,
\begin{equation}
b_{[i|j|k]}(f_1,f_2) =\frac{B_{[i|j|k]}(f_1,f_2)}{N_{(i|j|k)}(f_1,f_2)}
\label{th1}
\end{equation}
and its absolute value is upper bounded by one, i.e.,
\begin{equation}
\left|b_{[i|j|k]}(f_1,f_2)\right| \leq 1
\label{th2}
\end{equation}

To simplify notations, we will omit the dependence on the frequency in the following. The proof of equation \ref{th1} is fairly immediate. By inserting equation \ref{bicdef:app} in \ref{abicdef:app}, we get
\begin{equation}
b_{[i|j|k]} = \frac{B_{ijk}}{N_{(i|j|k)}} - \frac{B_{kji}}{N_{(k|j|i)}}
\end{equation}
and, since $N_{(i|j|k)}=N_{(k|j|i)}$ by construction, it follows that
\begin{equation}
b_{[i|j|k]} = \frac{B_{ijk}-B_{kji}}{N_{(i|j|k)}} =\frac{B_{[i|j|k]}}{N_{(i|j|k)}}
\end{equation}
which is what we wanted to show.

To demonstrate equation \ref{th2}, we will exploit the fact that the normalization factors are positive and real-valued, and thus they can be pulled out from the absolute value, i.e.
\begin{equation}
\left|b_{[i|j|k]}\right| = \frac{\left|B_{[i|j|k]}\right|}{N_{(i|j|k)}} = \frac{\left|B_{ijk}-B_{kji}\right|}{N_{ijk}+N_{kji}}
\end{equation}
It follows from the triangle inequality that 
\begin{equation}
\left|B_{ijk}-B_{kji}\right|\leq\left|B_{ijk}\right|+\left|B_{kji}\right|
\end{equation}
In addition, since $\left|B_{ijk}\right| \leq N_{ijk}$ and $\left|B_{kji}\right| \leq N_{kji}$, which follow from equation \ref{uninormbound}, we have
\begin{equation}
\left|b_{[i|j|k]}\right|  \leq \frac{\left|B_{ijk}\right|+\left|B_{kji}\right|}{N_{ijk}+N_{kji}} \leq \frac{N_{ijk}+N_{kji}}{N_{ijk}+N_{kji}} = 1
\end{equation}
which proves our assertion.

\begin{acknowledgments}
This work was supported by the Italian Ministry of Education, University and Research (PRIN 2010-2011 n. 2010SH7H3F\_006 ``Functional connectivity and neuroplasticity in physiological and pathological aging''), by the Fraunhofer Society of Germany, by grants from the EU (ERC-2010-AdG-269716), the DFG (SFB 936/A3/Z3), and the BMBF (031A130, 161A130). This project has received funding from the European Union's Horizon 2020 research and innovation programme under grant agreement No 686865. The content reflects only the author's view and the European Commission is not responsible for the content.
\end{acknowledgments}


%

\end{document}